\documentclass[fleqn,usenatbib,useAMS]{mnras}

\usepackage{hyperref}
\hypersetup{%
    colorlinks=true, 
    citecolor=blue} 

\usepackage[utf8]{inputenc}
\usepackage{upgreek}
\usepackage{amssymb}
\usepackage{color}
\usepackage[dvipsnames]{xcolor}
\usepackage{graphicx}	
\usepackage{supertabular}
\usepackage{caption}
\usepackage[T1]{fontenc}
\usepackage{ae,aecompl}

\usepackage{color}
\usepackage[normalem]{ulem}
\usepackage{cancel} 
\definecolor{refkey}{rgb}{1,1,1}
\definecolor{labelkey}{rgb}{1,0,0}	

\usepackage{mathtools}
\usepackage{mathrsfs}
\usepackage{dutchcal}

\newcommand{\vect}[1]{\mathbfit{#1}}

\usepackage{amsmath}
\usepackage{empheq}
\usepackage[most]{tcolorbox}

\definecolor{webgreen}{rgb}{0,.5,0}
\definecolor{webbrown}{rgb}{.6,0,0}
\definecolor{purple}{rgb}{0.5,0,.5}

\definecolor{blazeorange}{rgb}{1.0, 0.4, 0.0}
\definecolor{seagreen}{rgb}{0.18, 0.55, 0.34}
\definecolor{rufous}{rgb}{0.66, 0.11, 0.03}
\definecolor{royalfuchsia}{rgb}{0.79, 0.17, 0.57}
\definecolor{scarlet}{rgb}{1.0, 0.13, 0.0}
\definecolor{royalpurple}{rgb}{0.47, 0.32, 0.66}

\newcommand{\red}[1]{\textcolor{red}{#1}}
\newcommand{\green}[1]{\textcolor{green}{#1}}
\newcommand{\blue}[1]{\textcolor{blue}{#1}}

\newcommand{\cyan}[1]{\textcolor{cyan}{#1}}

\begin{document}
\label{firstpage}
\pagerange{\pageref{firstpage}--\pageref{lastpage}}


\title[Relativistic Oblique Shock Reflection]{A Numerical Study of Relativistic Oblique Shock Reflection}

\author[P. Bera et al.]{Prasanta Bera$^{1}$ \thanks{E-mail:pbera.phy@gmail.com}, Jonathan Granot$^{1,2,3}$, Michael Rabinovich$^{1,2}$, Paz Beniamini$^{1,2,3}$\\
$^{1}$Astrophysics Research Center of the Open University (ARCO), The Open University of Israel, P.O Box 808, Ra’anana 4353701, Israel\\
$^{2}$Department of Natural Sciences, The Open University of Israel, P.O Box 808, Ra’anana 4353701, Israel \\
$^{3}$Department of Physics, The George Washington University, 725 21st Street NW, Washington, DC 20052, USA
}
\pubyear{0000}
\date{Accepted XXX. Received YYY; in original form ZZZ}

\maketitle


\begin{abstract}
Shocks are ubiquitous in astrophysical sources, many of which involve relativistic bulk motions, leading to the formation of relativistic shocks. Such relativistic shocks have so far been studied mainly in one dimension, for simplicity, but the complex nature of the relevant astrophysical flows often requires higher dimensional studies. Here we study the two-dimensional problem of the reflection of a planer shock off of a wall for a general incidence angle and a cold unshocked medium. We use primarily relativistic hydrodynamic numerical simulations, and elaborately compare the results to an analytic treatment. The simulations are performed both in the rest frame S of the unshocked fluid, where the dimensionless proper speed of the singly shocked fluid is $u_1=\Gamma_1\beta_1$ and the shock incidence angle is $\alpha_1$, and in the rest frame S$^\prime$ of the point P of intersection of the incident shock and the wall for regular reflection (RR). Good agreement is obtained between the simulations in these two frames and with the analytic solution. The establishment of a steady flow in frame S$^\prime$ is explored, along with the transition between the strong and weak shock RR solutions. The transition line between RR and Mach reflection (MR) is studied numerically in the $u_1$\,-\,$\alpha_1$ plane and found to coincide with the analytic detachment/sonic line. 
The flow properties along the sonic line are investigated in detail focusing on how they vary between the Newtonian and relativistic limits.

\end{abstract}

\begin{keywords} 
shock waves -- relativistic processes -- methods: numerical -- hydrodynamics
\end{keywords}

\section{Introduction}

A steady single-phase subsonic inviscid flow maintains a smooth variation over different locations, excluding interfaces or boundaries. However, supersonic fluid velocities (with relative speeds between different parts of the fluid that exceed the sound speed) may form a discontinuity in matter density, pressure and velocity, which is termed a \textit{shock}. 

The location of the discontinuity (i.e., the shock) generally moves in space. 
The fluid crosses the shock from the upstream region to the downstream region and in the process, its density, pressure and specific entropy increase \citep[see e.g.,][]{Landau+Lifshitz1987,Thorne+Blandford2017}. Rankine–Hugoniot conditions specify the relationship between the fluid variables across the discontinuity \citep{Rankin1870, Hugoniot1887}. 
In the rest frame of the upstream fluid, the shock front moves supersonically, and the downstream shocked fluid carries nonzero momentum, kinetic energy and thermal energy. Shocks are very abundant in terrestrial and astrophysical fluids in supersonic motion.

In terrestrial phenomenon, the motion of a fluid (e.g. air, water) can attain a speed larger than the respective sound speed in the medium and this can form a shock. The head-on interaction of this discontinuity with a rigid wall, produces a reflected shock with a subsonic downstream region. In the case of an oblique incidence, the strength of the incident shock and the angle of the incidence determine the characteristics of the reflection. When the reflected shock and the incident shock intersect at a reflection point P on the wall, it is said to be regular reflection (RR). Otherwise, it is considered to be irregular reflection (IR), the most common configuration of which is called a Mach reflection (MR). In the case of MR, there exists a triple point ahead of the wall, where three lines intersect: the incident shock, the reflected shock, and a Mach stem \citep{vonNeuman1963,Courant+Friedrichs1948,Chester1954,Hornung1986,Olim+Dewey1992,Rosa+1992,Tabak+Rosales1994}. For large values of the shock incidence angle (defined as the angle between the shock and the wall), only IR/MR is possible, whereas for small incidence angle values only RR is possible.   

Shock reflection of non-relativistic oblique shocks was investigated in different experimental setups \citep{Heilig1969,Itoh+1981,Henderson+Gray1981} and numerical studies \citep{Mignone+2007,Gvozdeva+Chulyunin2015,wu2019study}. One of the main purposes of these studies was to pursue the transition criteria from RR to MR, and vice versa. Some theoretical criteria for this transition are known in the literature \citep[see, e.g.][]{vonNeuman1963,hornung1979transition,benDor1987}. For our purposes in this paper, the important criterion is the sonic criterion, which is discussed in detail below.

In an astrophysical environment, a fluid element can achieve a speed close to the speed of light, $c$, and form relativistic shocks,  
capable of generating significant radiation \citep{Blandford+McKee1976}.  The microscopic properties of the fluid are affected by relativistic thermal particle motions as
reflected in the equation of state \citep{Taub1948,Thorne1973}. Relativistic bulk motions have observational implications such as relativistic 
beaming effects \citep{Rees1966,Gold1969}. The strength (i.e. the Lorentz factor) of relativistic shocks may be inferred from the modeling of astrophysical objects, such as gamma-ray burst afterglows \citep{Sari1997,Rees+Meszaros1998}. 

Shocks play an important role in various astrophysical scenarios, such as: 
i) accretion by compact object \citep{Salpeter1964} ii) free-falling accretion onto the surface of a star, iii) interaction of stellar wind with the interstellar medium, iv) high-velocity ejecta from explosive transients, e.g. a nova, supernova or magnetar giant flare, v) in relativistic jets or outflows, such as gamma-ray bursts (GRBs), micro-quasars, active galactic nuclei (AGN), tidal disruption events, fast radio bursts or pulsar wind nebulae (PWNe), where shocks can form either due to collisions between different parts of the outflow (internal shocks) or due to its interaction with the ambient medium (external shocks). These astrophysical sources form some regions with very high internal energy density and the shocks accelerate both thermal and non-thermal electrons that produce bright radiation. Therefore the shock dynamics play a significant role in generating the observable radiation from many astrophysical sources.

Such astrophysical shocks may experience reflection by an obstacle. Some possible examples are: i) reflection of a supernova shock by the companion star in a binary stellar system \citep{Istomin+Soloviev2008}, ii) reflection of a GRB afterglow shock \citep{Lamberts+Daigne2018}, iii) reflection of shock formed at the magnetosphere by the stellar surface of a neutron star or the Sun, iv) reflection of a collimation shock at the jet-cocoon interface
with a cocoon in the cylindrical phase \citep{Adamson+Nicholls1958,Norman+1982}. To understand the underlying physics we can build a theoretical model relating the flow dynamics to the expected observable emission signatures. We follow a simplified approach to study the fluid dynamics of shock reflection in relativistic and non-relativistic regimes. We consider a perfectly reflecting wall as the reflector of the incident shocks.

In particular, in this work we numerically study the reflection of an incident oblique shock having Newtonian up to relativistic speeds, at different incidence angles. From direct relativistic hydrodynamic numerical simulation, we identify the characteristics of the reflected shock and find the criteria of RR. We compare our numerical results to analytic results derived in a companion paper 
\citep[][hereafter GR23]{GR23}
 
Initially, in \S\,\ref{sec:phy_setup}, we describe the physical setup of the numerical experiments. The underlying basic mathematical formulation is presented in \S\,\ref{sec:method}. Our results are presented in \S\,\ref{sec:results}
and the conclusions are discussed in \S\,\ref{sec:conclusions}.

\section{Physical setup}\label{sec:phy_setup}
\subsection{Lab frame S \& steady-sate frame S$^\prime$}\label{sec:SSprime}

\begin{figure}
 \includegraphics[width=0.98\columnwidth]{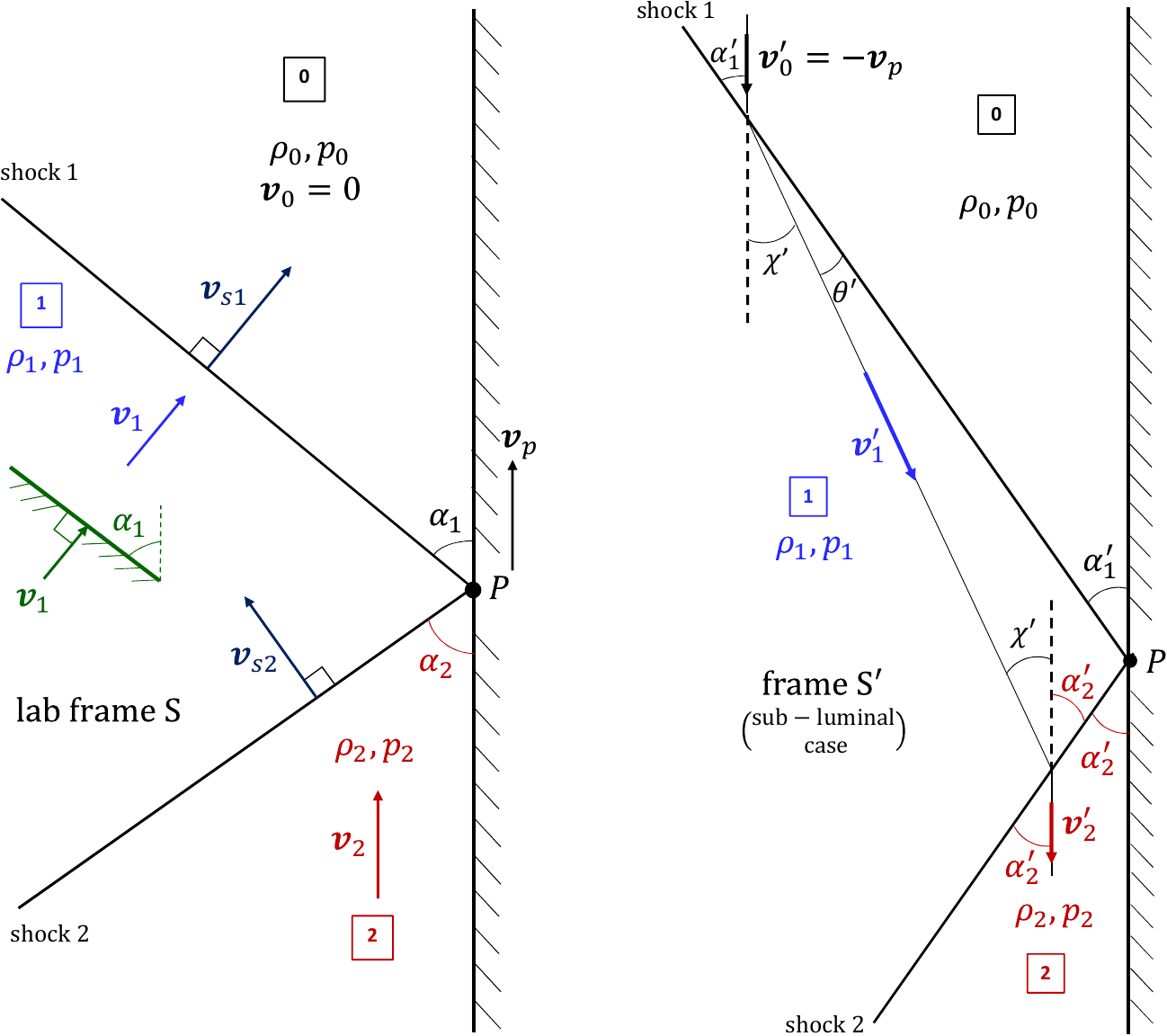}
    \caption{Schematic diagram of our setup for the shock reflection problem for RR, showing the location of the discontinuities. \textbf{\textit{Left}}: In the 
    lab frame S
    the unshocked cold fluid (region 0) is at rest and a piston moving at speed $v_1$ at an angle of $\alpha_1$ relative to a wall drives a shock ($s1$) into it (the shock front moving at speed $v_{s1}$) creating a singly shocked fluid region 1. The shock $s1$ hits the wall producing a reflected shock ($s2$) with a shock front moving at speed $v_{s2}$, and a doubly-shocked fluid region 2, whose velocity $v_2$ is parallel to the wall. The point $P$ where the two shocks intersect at the wall moves along the wall at a speed $v_p=v_{s1}/\sin\alpha_1=v_{s2}/\sin\alpha_2$. 
    \textbf{\textit{Right}}: In the rest frame S$^\prime$ of point P the flow is steady, and the fluid velocity in regions 0 and 2 is parallel to the wall. This rest frame exists only in the sub-luminal regime where $v_p<c\Leftrightarrow u_{s1}<\tan\alpha_1$ ($u_{s1}$ being the proper speed of shock s1).     }	
    \label{fig1_schemaic}
\end{figure}

The shock reflection is studied in two different frames of reference: i) the lab-frame S, where the unshocked region 0 is at rest, and ii) the moving frame S$^\prime$, where the flow is steady for RR (Figure~\ref{fig1_schemaic}). 

Initially, we set up the problem in the lab frame S. There are two shocks labeled 1 (incident shock) and 2 (reflected shock), which divide the flow into three regions, labeled 0, 1 and 2,  corresponding to the number of times the fluid in each region was shocked. The unshocked region 0 is adjacent to a perfectly reflecting static wall and considered to be at rest in frame S (velocity $v_0=0$) and cold (pressure $p_0/\rho_0c^2\ll\min(1,u_1^2)$, where $\rho_0$ is its proper rest-mass density and $u_1=\Gamma_1\beta_1$ is the dimensionless proper speed of region 1). In frame S, the incident shock (`shock 1') moves with a velocity $v_{s1}$ along its normal and makes an angle $\alpha_1$ with respect to the wall. It can be thought of as generated by a piston moving at velocity $v_1<v_{ s1}$ and driving a shock with a velocity $v_{ s1}$ (see Fig.~\ref{fig1_schemaic}). The proper rest-mass density $\rho_1$ and pressure $p_1$ in region 1 are determined by the jump conditions of shock 1. The collision of the incident shock 1 with the wall creates a reflected `shock 2', with a velocity $v_{s2}$ along its normal, and making an angle $\alpha_2$ with respect to the wall. A post-shock region 2 forms between the wall and shock 2, with proper rest-mass density $\rho_2$, pressure $p_2$ and velocity $v_2$.  

As the incident shock 1 is oblique, $\alpha_1>0$, it intersects the wall at a point P, which moves along the wall at a velocity $\boldsymbol{v}_p$,whose magnitude is given by 
\begin{equation}\label{eq_v_P}
v_p=\frac{v_{s1}}{\sin\alpha_1}=\frac{v_{s2}}{\sin\alpha_2}\ .
\end{equation}
In the case of a semi-infinite steady oblique incident shock, the incident and reflected shocks move in a self-similar pattern with respect to point P. In the sub-luminal regime that corresponds to $\beta_p=v_p/c<1$, one can transform (through a boost at $-\boldsymbol{v}_p$) to a rest frame S', in which for RR, point P is at rest and the flow is steady.  
In frame S$^\prime$, region~0 moves with a velocity $\boldsymbol{v}'_0=-\boldsymbol{v}_p$. Similarly, the velocity of region~1 and the incident shock are Lorentz-boosted by $-\boldsymbol{v}_p$ from the S-frame values. The proper rest-mass densities ($\rho_0,\rho_1$) and pressures ($p_0,p_1$) are invariant. 
The angles formed by the incident and reflected shocks with the wall in S$^\prime$ are also Lorentz boosted, i.e., 
\begin{equation}\label{eq:engle_trans}
\tan\alpha'_{i} = \frac{\tan\alpha_{i}}{\Gamma_p}\quad\quad(i=1,\,2)\ ,
\end{equation}
where $\Gamma_p=(1-\beta_p^2)^{-1/2}$. 

\subsection{The General Structure of the Parameter Space}\label{sec:par_space}

\begin{figure}
 \includegraphics[width=0.99\columnwidth]{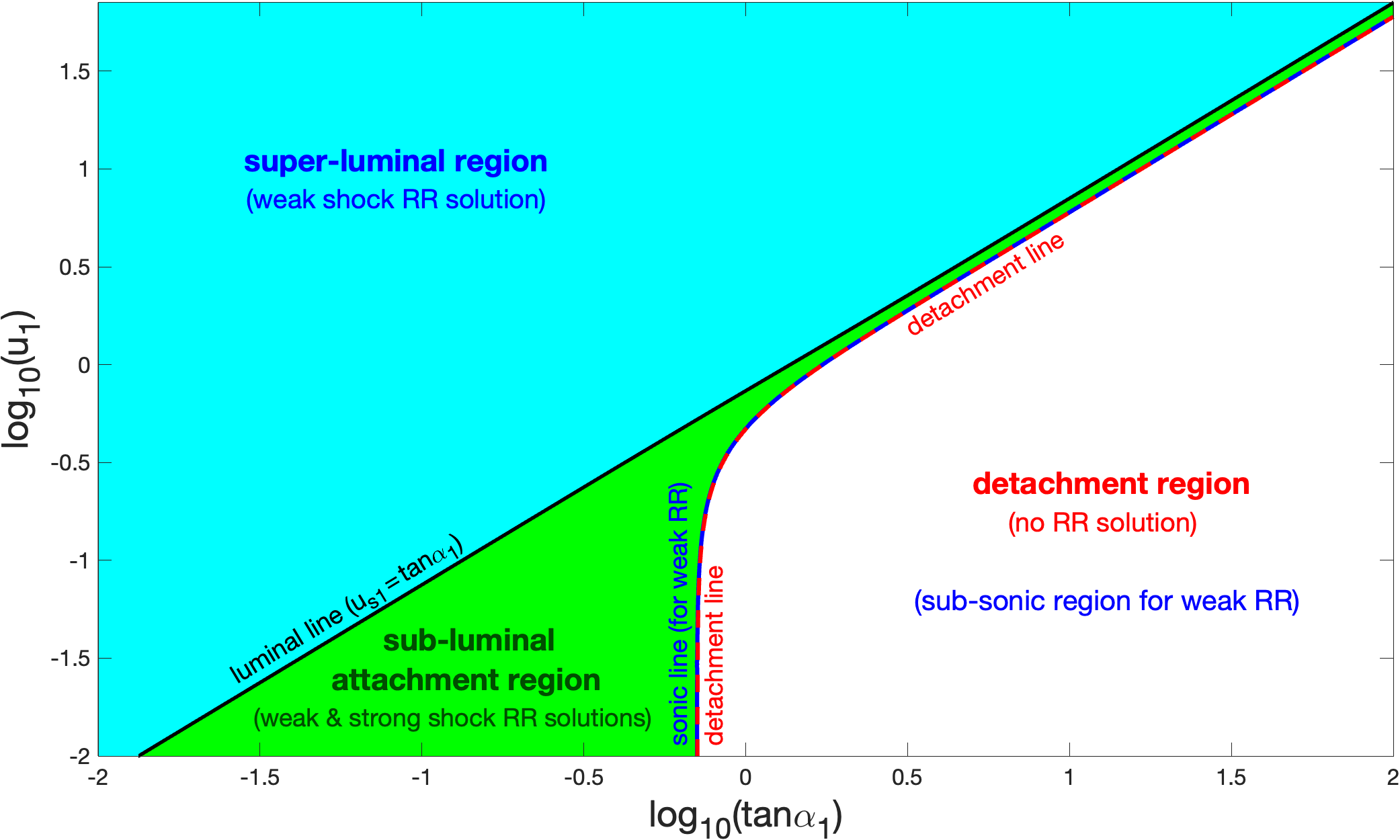}
    \caption{The different regions and critical lines in the $u_1$\,--\,$\alpha_1$ parameter space, shown in terms of $\log_{10}(u_1)$ versus $\log_{10}(\tan\alpha_1)$.
    The luminal line (\textit{in black}; $\beta_p=1\Leftrightarrow u_{s1}=\tan\alpha_1$) separates the super-luminal region (\textit{\cyan{cyan shading}}) and the sub-luminal 
    attachment region (\textit{\green{green shading}}), which is in turn separated from the detached region (\textit{in white}, where there is no regular reflection -- RR)
    by the detachment line (\textit{\red{in dashed red}}), which almost coincides with the sonic line for the weak RR solution
    (\textit{\blue{in dashed blue}}; $\beta'_{2,w}=\beta_{c_s,2,w}$, see GR23 for details).}
    \label{fig:critical_lines}
\end{figure}

We study oblique reflected shocks with different incidence angles, $\alpha_1$, and different proper velocities of the incident fluid, $u_1=\Gamma_1\beta_1$.
Figure~\ref{fig:critical_lines} shows the analytic expectation (as derived in GR23) for the different regions in the $u_1$\,--\,$\alpha_1$ parameter space, and the critical lines that separate between them. This is displayed by showing $\log_{10}(u_1)$ in the $y$-axis versus 
$\log_{10}(\tan\alpha_1)$ 
in the $x$-axis. The luminal line (\textit{in black}; defined by the condition $v_p=c$ or, equivalently, $u_{s1}=\tan\alpha_1$) separates the super-luminal region (in \textit{cyan shading}) and the sub-luminal regions. The sonic line for the weak shock RR solution (\textit{in \blue{dashed blue}}; defined by $\beta'_{2,w}=\beta_{c_s,2,w}$ where the subscript `w' stands for the weak shock RR solution) is found (GR23) to almost coincide with the detachment line (\textit{in \red{dashed red}}), which bounds the region with RR solutions.
We shall therefore not make the distinction between them here, and refer mainly to the sonic line.
The sonic line always lies in the sub-luminal region\footnote{This is since it corresponds to $\beta_p=(\beta_{2,w}+\beta_{c_s,2,w})/(1+\beta_{2,w}\beta_{c_s,2,w})<1$, i.e. the sonic condition implies that $v_p$ is equal to the lab-frame speed of a sound wave moving in region 2 parallel to the wall, which must therefore be less than $c$.}  and separates between the sub-sonic (or detachment) region (\textit{in white}), where there is no RR solution (but instead, there is IR -- a more complicated type of shock reflection, such as MR), and the super-sonic (or attachment) regions. The region between these two critical lines -- the sub-luminal super-sonic (or attachment) region is marked in \textit{green shading}. 

In the following, we find numerically that, as expected analytically, the sonic line bounds the region of RR. 
There can in principle also be a dual region where both MR and RR are possible for the same $(u_1,\alpha_1)$ values. Such a dual region borders the sonic line on the super-sonic side but is located well within the sub-luminal region. The fact that we do not find such a dual region might be since the RR weak shock solution is a more stable attractor solution, such that the MR solution is not found in the simulations, similar to the RR strong shock solution that is discussed in \S\,\ref{subsec_regular_reflection}.

Therefore, the sonic line is of particular physical importance. While it was extensively studied in the Newtonian regime, it was not studied before in the relativistic regime. We study it here in detail, stressing the differences between the Newtonian and relativistic regimes, and how the system transitions between these two limits.

\begin{figure}
	\includegraphics[width=0.97\columnwidth]{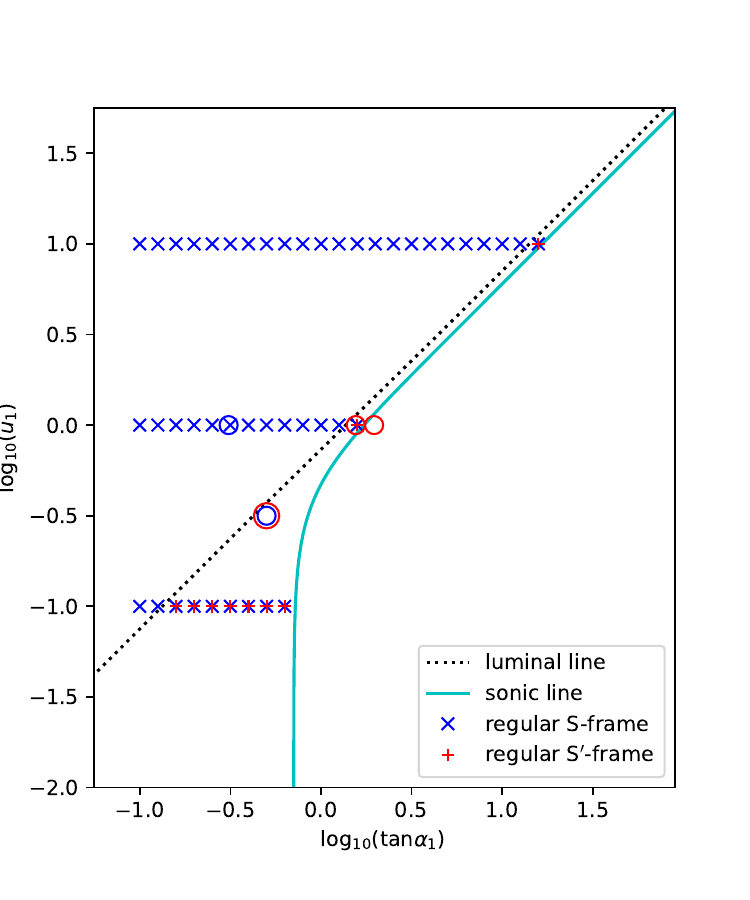}
    \caption{Scatter marks represent the parameter space coverage of the numerical calculations presented in section~\ref{sec:results}. The \textcolor{blue}{blue}, \textcolor{magenta}{magenta} and \textcolor{red}{red} circles correspond to the snapshots of RR in S and S$^\prime$ and MR in S$^\prime$ (section~\ref{sec:results_u1alpha1}) respectively. For a small inclination angle, the S-frame captures RR effectively. The frame S$^\prime$ is applicable for the incident angle higher than the luminal boundary.}	
    \label{fig:numeric_runs}
\end{figure}
We study shock reflection in the ($u_1-\alpha_1$) parameter space. We use the above-mentioned reference frames S and S$^\prime$. Figure~\ref{fig:numeric_runs} shows the points for which we performed special relativistic hydrodynamic numerical simulations
(section~\ref{sec:method}) to obtain the outcome of the shock reflection by a wall. 

\section{Numerical Method}\label{sec:method}
The conservation equations for total mass, momentum and energy in the special theory of relativity may be written as: 
\begin{empheq}{align} 
    \partial_\mu(\rho u^\mu)&=\frac{\partial (\rho\Gamma)}{\partial t}+\nabla\cdot(\rho\vect{u})=0\ ,\label{eq_mass_con}\\
    \partial_\nu T^{i\nu}&=\frac{\partial (w\Gamma\vect{u})}{\partial t}+\nabla\cdot(w\vect{u}\vect{u}+p\boldsymbol{I})=0\ ,\label{eq_mom_con}\\
    \partial_\nu T^{0\nu}&=\frac{\partial (w\Gamma^2-p)}{\partial t}+\nabla\cdot(w\Gamma\vect{u})=0\ ,\label{eq_energy_con}
\end{empheq}
where $T^{\mu\nu}=wu^\mu u^\nu+p\eta^{\mu\nu}$ is the stress-energy tensor, $\eta^{\mu\nu}$ is the Minkowski metric, $u^\mu$ is the 4-velocity, $\mathbf{u}=\Gamma\boldsymbol{\beta}=u\hat{\boldsymbol{u}}$ is the proper velocity of the fluid, $\boldsymbol{\beta}=\boldsymbol{v}/c=\beta\hat{\boldsymbol{\beta}}$, $\Gamma=(1-\beta^2)^{-1/2}$ is the Lorentz factor, $\rho$ is the proper rest mass density, $p$ is the pressure, $w=e+p=\rho c^2+e_{\rm int}+p$ is the proper enthalpy density, and $e$  ($e_{\rm int}$) is the proper (internal) energy density.
Here $\frac{\partial}{\partial t}$, $\nabla$ and $\boldsymbol{I}$ are the time derivative, spatial derivative and the unit $3\times3$ matrix, respectively.

In the presence of a 1D shock, the fluid variables 
on its two sides (0,1: pre- \& post-shock regions) 
satisfy the following (Rankine–Hugoniot) jump conditions  conditions, 

\begin{align}
    \rho_0\Gamma_{0,s1}\beta_{0,s1} &= \rho_1\Gamma_{1,s1}\beta_{1,s1}\ \label{eq_shock_con_mass},\\
    w_0\Gamma_{0,s1}^2\beta_{0,s1}^2+p_0 &= w_1\Gamma_{1,s1}^2\beta_{1,s1}^2+p_1\ \label{eq_shock_con_mom},\\
    w_0\Gamma_{0,s1}^2\beta_{0,s1} &= w_1\Gamma_{1,s1}^2\beta_{1,s1}\ \label{eq_shock_con_eng},
\end{align}
where quantities relating to the upstream (pre-shock) region and the downstream (post-shock) region are denoted by subscripts 0 and 1, respectively.
These jump conditions may be obtained by equating the fluxes of matter, momentum and energy on the two sides of the shock, in the frame where the shock front is at rest and the fluid velocities are normal to it. The velocities $\beta_{0,s1}$ and $\beta_{1,s1}$ are those of regions 0 and 1 in the shock 1 rest-frame, while $\Gamma_{i,s1}=(1-\beta_{i,s1}^2)^{-1/2}$ are the corresponding Lorentz factors. The pressure, density and normal component of velocity are discontinuous across the shock. 

To solve the above set of fluid equations and to obtain the values of downstream fluid for the given upstream values we need to provide the equation of state (EoS). To capture the relativistic and non-relativistic regimes we consider the following equation \citep{Mignone+McKinney2007}:
\begin{align}
    \left(h-\Theta\right)\left(h-4\Theta\right)=1 \label{eq_EoS_Taub}\ ,
\end{align}
where $\Theta=p/\rho c^2$ and the enthalpy per unit rest energy $h$ and the effective adiabatic index $\hat\gamma$ are give by 
\begin{eqnarray}
h&=&1+\frac{\hat{\gamma}\Theta}{\hat{\gamma}-1}=\frac {5}{2}\Theta+\sqrt{1+\frac{9}{4}\Theta^2}\ ,
\\
\hat{\gamma}&=&
\frac{\frac{\partial h}{\partial \Theta}}{\frac{\partial h}{\partial \Theta}-1}=\frac{1}{6}\left(8-3\Theta+\sqrt{4+9\Theta^2}\right)\ .\quad\quad
\end{eqnarray}
This EoS satisfies the \citet{Taub1948} inequality of relativistic matter. The corresponding dimensionless sound speed, $\beta_{c_s}=c_s/c$, is given by \citep{Ryu+06}
\begin{align}
\beta_{c_s}^2=\frac{\partial p}{\partial e}=\frac{\Theta}{h}\frac{\frac{\partial h}{\partial \Theta}}{\frac{\partial h}{\partial \Theta}-1}=\frac{3\Theta^2+5\Theta\sqrt{\Theta^2+4/9}}{12\Theta^2+2+12\Theta\sqrt{\Theta^2+4/9}}\ .
\end{align}
Here we aim to find the values of the downstream quantities from the direct hydrodynamic simulations.
To this end we study the shock reflection in frames S \& S' described above.

\subsection{Numerical setup in frame S}\label{sec_numerical_setup_S}
In order to study the shock reflection problem described in \S\,\ref{sec:phy_setup} in frame S, 
we 
set up the incident shock~s1 
in this frame by prescribing regions 0 and 1 in the computation domain. We 
then numerically solve the time evolution of the computation domain, identifying the different regions and the critical lines that separate them. In particular,
we track the formation of the reflected shock~s2 and the doubly shocked region~2 as it is described in \S\,\ref{sec:SSprime}. To reduce the artifacts from the numerical scheme, we avoid ultra-low values of pressure in region~0 and choose a moderately low value 
of $\Theta_0=p_0/\rho_0 c^2\sim10^{-9}$ to represent a cold medium of region~0. The fluid  is at rest in region~0 ($\boldsymbol{v}_0=0$) while the velocity of region~1 is $\boldsymbol{v}_1$. The pressure ($p_1$) and proper rest-mass density ($\rho_1$) in region~1, as well as the velocity of shock s1 along its normal in frame~S ($\boldsymbol{v}_{s1}$) are obtained by solving the shock jump conditions (equations~(\ref{eq_shock_con_mass})-(\ref{eq_shock_con_eng})).  

We use the \textsc{pluto} code \citep{Mignone+2007} to solve the hydrodynamic equations~(\ref{eq_mass_con})-(\ref{eq_energy_con}) in a fixed linear spaced grid. We use the piece-wise parabolic reconstruction scheme, with the second-order Runge-Kutta time integration and HLLC Reimann solver. We choose the initial shock location along the diagonal of the computational domain connecting the top-left and the bottom-right. We consider a few hundreds to thousands of grid points in each side of the two-dimensional computational domain maintaining near near-equal aspect ratio of grid-spacing. The presence of the wall at the r.h.s. boundary is obtained by implementing reflecting boundary conditions. The region~1 inflow boundary conditions are implemented at the left and at the bottom of the computation domain. The top of the computation domain is maintained with a free outflow condition.

In the lab frame S, region~1 increases its area as the point of contact P moves along the wall. The shape of the post-shock region i.e. the angle $\alpha_2$ is not known a priori. To start the simulation we use the inflow condition from the left 
and lower boundaries.
As the 
doubly-shocked region 2 develops and it interferes with this fixed inflow condition along the lower boundary. To prevent the impact of the boundary conditions on the results we focus our analysis on a region sufficiently close to the point of contact P, such that it is not affected by the lower boundary condition.

\subsection{Numerical setup in frame S$^\prime$}\label{sec_numerical_setup_Sprime}

By construction, the shock~s1 remains static in frame~S$^\prime$ and the region~0 is confined between the shock~s1 and the wall which makes an angle $\alpha_1^\prime$ at P, given by $\tan\alpha_1^\prime=\Gamma_p^{-1}\tan\alpha_1$. Region~0 has proper rest mass density $\rho_0$, pressure $p_0$ and velocity $\boldsymbol{v}'_0=-\boldsymbol{v}_p$. Region~1 has proper rest mass density $\rho_1$, pressure $p_1$ and the velocity is given by a Lorentz boost by
$-\boldsymbol{v}_p$ from the frame S value
$\textbf{\textit{v}}_1=\beta_1 c(-\cos\alpha_1,\,\sin\alpha_1)$,
\begin{equation}
\boldsymbol{v}'_1=\frac{\left[-v_1\cos\alpha_1,\,\Gamma_p(v_1\sin\alpha_1-v_p)\right]}{\Gamma_p(1-\beta_p\beta_1\sin\alpha_1)}\ .
\end{equation}

In the \textsc{pluto} setup we start the numerical 
simulation with the fluids in regions~0 and 1. The top and left edges of the computational domain maintain inflow boundary conditions of regions~0 and 1, respectively, while the bottom edge maintains a free outflow boundary condition.
As the incident shock s1 impacts the wall, it forms the reflected shock~s2 and the doubly-shocked region 2 develops. 

We also test the dynamical stability of the RR strong shock solution, by adding the corresponding algebraic solution for region~2 to the initial conditions of the simulation (as this solution is unstable and does not otherwise develop naturally in the numerical simulation). In this case the point of transition between the inflow boundary conditions of regions~0 and 1 is no longer at the top left corner, but is instead located at a fixed point along the top edge of the computational domain.

We consider the evolution in frame S$^\prime$ in the vicinity of point P, as much as possible. The advantage of this frame is that the incident and reflected shocks are static, and the flow is steady for RR. 
However, since $\beta_p=\beta_{1s}/\sin\alpha_1$, for high incident shock speeds $\beta_{1s}$ and/or small incidence angles $\alpha_1$,
the velocity of point P might rise above the speed of light, and in this super-luminal regime frame S' does not exist.
We start a numerical evolution of fluid in region~1 and gradually it develops the region~2. 
As the post-shock region (region~2) develops, we select a region away from the boundary and find the location of the discontinuity.

When performing simulations in frame S' we inject fluid into region 0 and 1 with velocities $\boldsymbol{v}'_0$ and $\boldsymbol{v}'_1$, from the upper and left boundaries of the simulation box, respectively.

We have the freedom to choose arbitrary time units, $t_{\rm unit}$ in S and $t^\prime_{\rm unit}$ in S$^\prime$ (corresponding to length units $l_{\rm unit}=ct_{\rm unit}$ and $l^\prime_{\rm unit}=ct^\prime_{\rm unit}$), to design the frame for the direct numerical study. 
In frame S we measure the simulation time in units of shock crossing time, $t_{p}=L_y/v_p$, where $L_y$ is the simulation box size along the wall. In frame S$^\prime$ we use as our time unit the sound crossing time of the doubly shocked region 2, $t^\prime_{sc}=L'/c_{s,2}$, where $L'$ is its length along the wall and $c_{s,2}$ is the sound speed in region~2. 

\section{Results}\label{sec:results}

\subsection{The $u_1$\,--\,$\alpha_1$ parameter space}
\label{sec:results_u1alpha1}

Here we summarize the results obtained from our numerical simulations of shock reflection for different proper speeds $u_1$ of the singly shocked region~1 and different incidence angles $\alpha_1$. 
For a given $u_1$, RR is expected for small enough values $\alpha_1$ (see Figure~\ref{fig:critical_lines}).

\subsubsection{Regular Reflection (RR)}\label{subsec_regular_reflection}

\begin{figure}
\includegraphics[width=0.988\columnwidth]{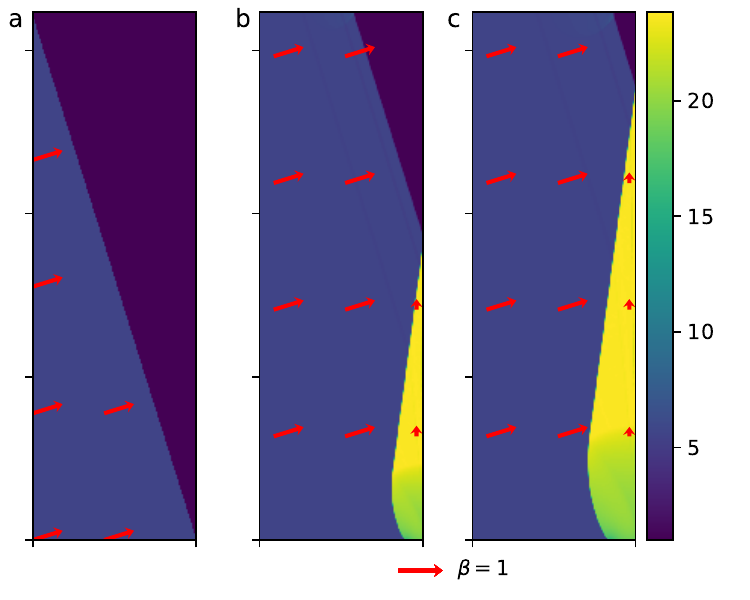}
    \caption{Snapshots from a numerical simulation of shock reflection in the lab frame S for RR, at different times: (a) $t/t_p=0.0$, (b) $t/t_p=0.57$, (c) $t/t_p=0.86$, where $t_{p}=L_y/v_p$ is the box crossing time of point P (using equal aspect ratio). The computation domain of lengths ratio $L_y:L_x=3.23:1$, where the reflecting wall is along its right boundary while the incident shock s1 is initially along its top-left to bottom-right diagonal.
    The unit of length is arbitrary. The incident shock s1 leaves the computation domain at $t=t_p$. The color scale indicates the fluid's proper rest-mass density while the red arrows show its velocity vector. 
    This simulation is initialized with
    $\alpha_1=0.3$ and $u_1=1$, where the latter implies $u_{1s}=1.37$ and $\rho_1/\rho_0=4\Gamma_1=4\sqrt{2}$ for $p_0\ll\rho_0c^2u_1^2$. 
    Panel (a) shows the initial conditions. Other snapshots (panels~(b),\,(c)) indicate the gradual growth of the doubly shocked region~2 with a higher density. The reflected shock forms an angle $\alpha_2=0.137$ with the wall. 
    Region 2 remains uniform unless it is affected by the lower boundary condition.}	
    \label{fig_ShockWall_Lab_regular}
\end{figure}

\begin{figure}
	\includegraphics[width=0.988
 \columnwidth]{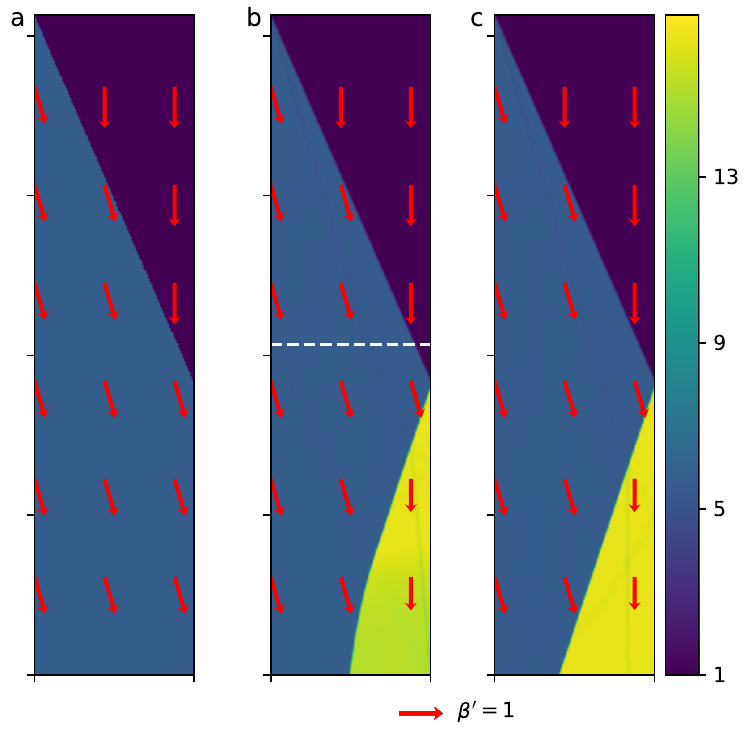}
\caption{Snapshots from a numerical simulation of shock reflection in the frame S' for RR, at different times:
(a) $t^\prime/t_{sc}^\prime=0$, (b) $t^\prime/t_{sc}^\prime=0.72$, (c) $t^\prime/t_{sc}^\prime=1.43$ (where $t_{sc}^\prime$ is defined in \S\,\ref{sec:SSprime}). 
    The height-to-width ratio of the computation domain is $L'_y/L'_x=4.13$. The doubly-shocked region~2 develops between the wall and the reflected shock s2, which forms an angle $\alpha'_2$ (with $\tan\alpha_2^\prime=0.33$) relative to the wall. Figure~\ref{fig_ShockWall_Mov_regular_sequences} shows zoomed-in snapshots (of the region below the dashed white line in panel (b)), more densely sampled in time to better illustrate the formation of region~2.
    }	
    \label{fig_ShockWall_Mov_regular}
\end{figure}

We have captured the time evolution of incident shock~s1 and the development of reflected shock s2 by performing numerical simulations in two different rest frames. In the lab frame S, both the incident and the reflected shocks move (\textit{left panel} of Figure~\ref{fig1_schemaic} and snapshots from the numerical study in Figure~\ref{fig_ShockWall_Lab_regular}). The same incident and reflected shocks remain steady at the moving frame $S^\prime$ (\textit{right panel} of Figure~\ref{fig1_schemaic} and snapshots from the numerical study in Figure~\ref{fig_ShockWall_Mov_regular}). 

In frame S we start the numerical simulation of an incident shock s1 with $u_1=1,\beta_1=1/\sqrt{2}\approx0.7071,\beta_{1s}=4\sqrt{2}/7\approx0.8081$,
$\rho_1/\rho_0=4\sqrt{2}\approx5.657$, 
$p_1=4/3$)
and an incidence angle $\alpha_1=0.3$ relative to the reflecting wall (at the bottom right corner of Figure~\ref{fig_ShockWall_Lab_regular}a). Therefore, initially there is no doubly-shocked region 2. The intersection point P of the incident shock s1 and the wall moves along the wall at a speed $v_p=\beta_{1s}/\sin\alpha_1$. We consider a computation box (a $568\times918$ grid in the $x$-$y$ plane) with its vertical length ($L_y$) along the wall being 3.23 times larger than the horizontal length ($L_x$). We measure the time in units of the shock crossing time i.e. $t_{p}=L_y/v_p$. As time evolves, a high-density doubly-shocked region~2 develops, between the wall and the reflected shock s2, which makes
an angle $\alpha_2=0.137$ relative to the wall (Figure~\ref{fig_ShockWall_Lab_regular}b~\&~\ref{fig_ShockWall_Lab_regular}c). The similarity between the Figures~\ref{fig_ShockWall_Lab_regular}b~\&~\ref{fig_ShockWall_Lab_regular}c indicates the self-similar 
nature of the flow with respect to the point P. 
The fluid in region~2 moves along the wall, relatively slowly, at a proper speed $u_2=0.247$ for $(u_1,\alpha_1)=(1,0.3)$.
The  inflow boundary condition at the lower boundary is unphysical within region~2, 
and its effects become more
significant for a higher value of $\alpha_1$.

In frame S' we initialized the numerical simulations as described in \S\,\ref{sec_numerical_setup_Sprime}. Figure~\ref{fig_ShockWall_Mov_regular} shows snapshots from such a simulation with $(u_1,\alpha_1)=(1,1)$ (or $\tan\alpha_1=1.557$), such that ($\rho_0=1$, $p_0=10^{-9}\rho_0c^2$) imply ($\beta_{0y}^\prime=-\beta_p=-0.907$, $\rho_1=5.657$, $p_1=1.333$, $\tan\alpha'_1=0.434$).
We start the numerical simulation with regions~0 and 1 (Figure~\ref{fig_ShockWall_Mov_regular}a) in the computational box (of $252\times1038$ grid points) with its vertical length along the wall ($L'_y$) being 4.13 times the horizontal length ($L'_x$). The 
doubly-shocked region~2 develops with time (Figure~\ref{fig_ShockWall_Mov_regular}b~\&~\ref{fig_ShockWall_Mov_regular}c) as the incident shock s1 remain static in this frame. The reflected shock s2 settles down at an angle $\alpha_2^\prime$ (with $\tan\alpha_2^\prime=0.33$) relative to the wall. The matter in region~2
($\rho_2/\rho_0=16.171$, $p_2/\rho_0 c^2=7.039$) moves along the wall with a proper speed $u^\prime_2=1.207$. 

\begin{figure}
	\includegraphics[width=0.988
 \columnwidth]{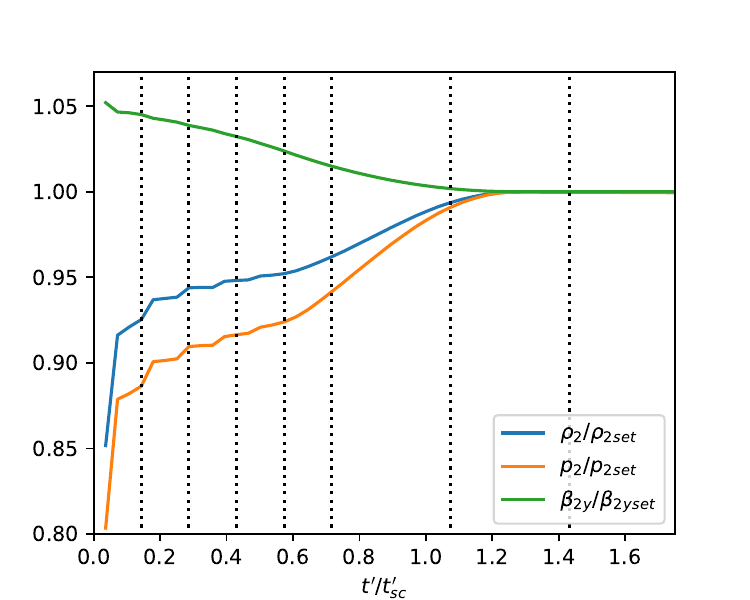}
    \caption{Average density ($\rho_2$), pressure ($p_2$) and velocity component along the wall ($\beta_{2y}$) of the doubly-shocked region~2, as it forms and settles down (to $\rho_{\rm 2,set}$, $p_{\rm 2,set}$,  and $\beta_{2y{\rm,set}}$, respectively) over about a sound crossing time ($t_{sc}^\prime$). The vertical dotted lines indicate the time stamps of the snapshots shown in Figure~\ref{fig_ShockWall_Mov_regular_sequences}.}	
    \label{fig_ShockWall_Mov_regular_ettlement_process}
\end{figure}

Figure~\ref{fig_ShockWall_Mov_regular_ettlement_process} shows that 
the doubly-shocked region~2 forms and settles down over a timescale close to its sound crossing time, $t^\prime_{sc}=L'/c_{s,2}$, where $L'$ is its length along the wall. We identify region~2 through its higher density relative to region~1.
Region~2 contains proper-density fluctuations 
of a few percent ($\lesssim5\%$) relative to the mean value. The non-uniformity in region~2 is due to the gradual transition at the boundary and the fluctuations. 

\begin{figure*}
	\includegraphics[width=1.988
 \columnwidth]{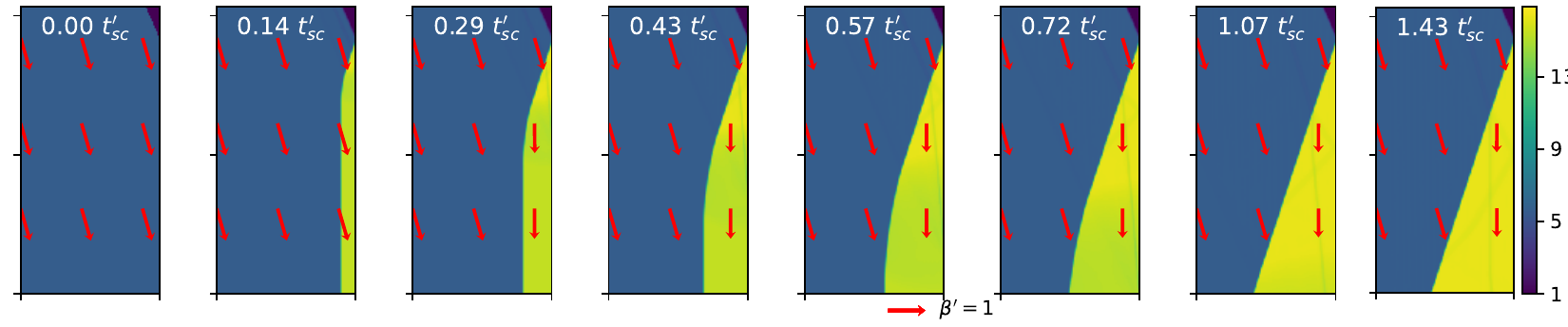}
    \caption{The gradual development of the doubly-shocked region~2 is shown in this sequence of snapshots in frame S', depicting 
    the lower half of the computational domain 
    from Figure~\ref{fig_ShockWall_Mov_regular} 
    at the times indicated by the vertical dotted line in Figure~\ref{fig_ShockWall_Mov_regular_ettlement_process}. }	
    \label{fig_ShockWall_Mov_regular_sequences}
\end{figure*}

Figure~\ref{fig_ShockWall_Mov_regular_sequences} shows the snapshots displaying the density and velocity of the doubly-shocked region~2 at different times. Frame S$^\prime$ is suitable to study the shock interaction for high enough $\alpha_1$ values, corresponding to the sub-luminal region.

\begin{figure}
	\includegraphics[width=0.988
 \columnwidth]{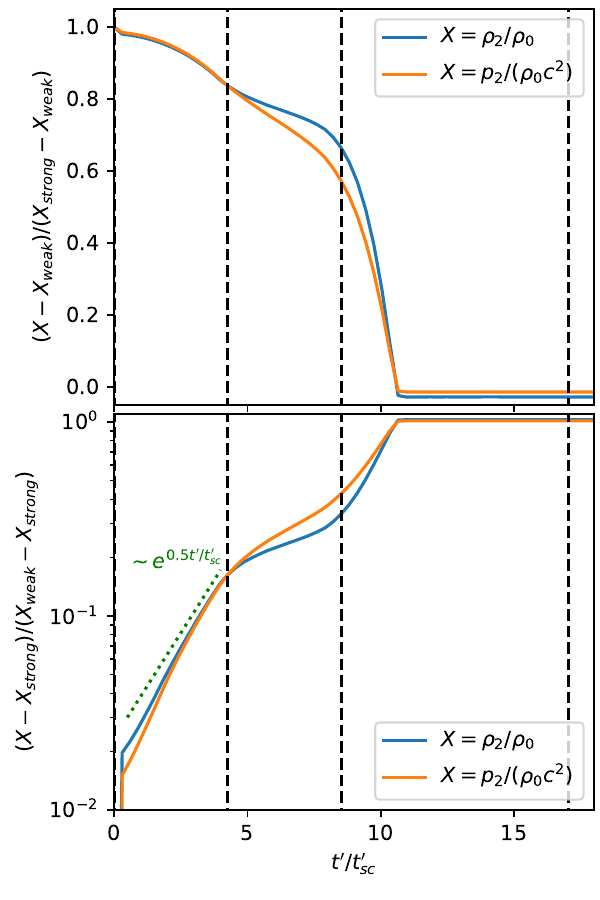}
    \caption{The evolution of the mean proper rest-mass density ($\rho_2$) and pressure ($p_2$) in the doubly-shocked region~2, shown in terms of their fractional deviations from their values for the `strong' ($\rho_{\rm strong}$, $p_{\rm strong}$) and `weak' ($\rho_{\rm weak}$, $p_{\rm weak}$) shock RR solutions.
    The simulation starts from the algebraic `strong' shock solution and the system     moves to the `weak' shock solution
    within several sound-crossing times of region 2 ($t_{\rm sc}^\prime$). The vertical dashed lines denote the times of the snapshots shown in Figure~\ref{fig_S2Wsnaps_rhoprs}.}
    \label{fig_S2Wtime_evolution}
\end{figure}

\begin{figure*}
	\includegraphics[width=1.988
 \columnwidth]{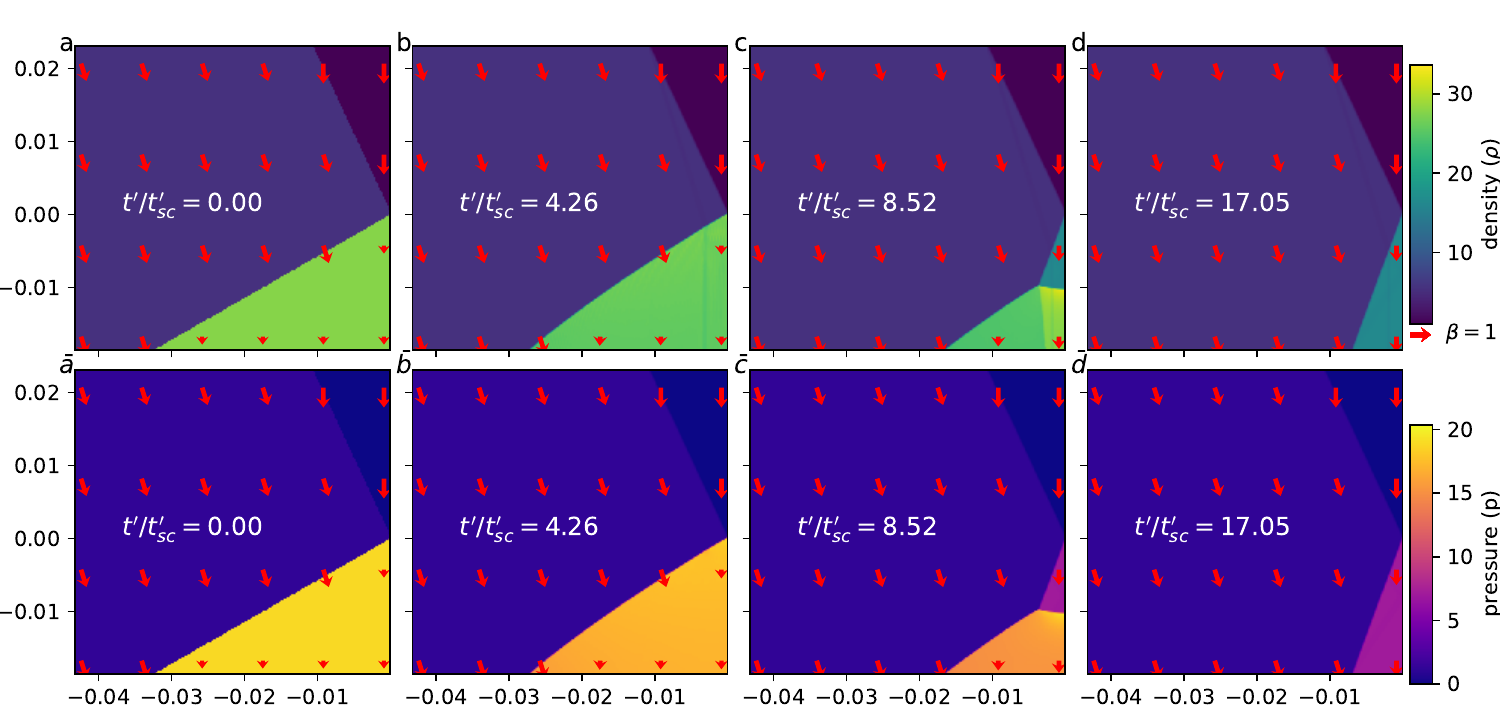}
    \caption{Snapshots of proper rest-mass density (\textit{top panels} a-d) and pressure (\textit{bottom panels} $\bar{\rm a}$-$\bar{\rm d}$) 
    from a simulation in frame S' starting from the algebraic strong shock RR solution (\textit{panels} a,\,$\bar{\rm a}$). The red arrows are velocity vectors, whose size indicates the fluid speed at their starting point.  
    This time sequence captures the evolution between the strong and weak shock solutions, in the initial linear phase (\textit{panels} b,\,$\bar{\rm b}$) and subsequent non-linear phase (\textit{panels} c,\,$\bar{\rm c}$). The system finally settles in the weak shock solution (\textit{panels} d,\,$\bar{\rm d}$).}	
    \label{fig_S2Wsnaps_rhoprs}
\end{figure*}

For RR,
our simulations (in frame S$^\prime$) that initially did not contain the doubly-shocked region~2, evolved to and settled at the `weak' shock RR solution.
From the algebraic solution, one may also obtain a `strong' shock RR solution (in the super-sonic sub-luminal region -- see GR23), corresponding to higher values of $\alpha'_2$ (and therefore $\alpha_2$), $\rho_2$ and $p_2$. 
To study the stability of this strong shock solution we numerically evolve the fluid variables starting from an initial configuration that also includes region~2 with properties corresponding to the algebraic strong shock solution (see Table~\ref{table_weak_strong}, Figures~\ref{fig_S2Wtime_evolution} and \ref{fig_S2Wsnaps_rhoprs}).

\begin{table}
\centering
\begin{tabular}{||c |c c| c c c||} 
\hline 
\multicolumn{3}{|c|}{Frame S}&\multicolumn{3}{|c|}{Frame S$^\prime$}\\
\hline\hline
$\rho_{0}$& \multicolumn{2}{|c|}{1}& $\rho_{0}$ & \multicolumn{2}{|c|}{1}\\
\hline
$\beta_{0y}$& \multicolumn{2}{|c|}{0}& $\beta_{0y}^\prime$ & \multicolumn{2}{|c|}{-$v_p$=-0.9555363}\\
\hline
$\alpha_{1}$& \multicolumn{2}{|c|}{1.0079245}& $\alpha_{1}^\prime$ & \multicolumn{2}{|c|}{0.43718}\\
\hline
$u_{1}$& \multicolumn{2}{|c|}{1}& $u_{1}^\prime$ & \multicolumn{2}{|c|}{1.795769}\\
\hline
$\rho_{1}$& \multicolumn{2}{|c|}{5.6568}& $\rho_{1}$ & \multicolumn{2}{|c|}{5.6568}\\
\hline
$p_{1}$& \multicolumn{2}{|c|}{1.33333}& $p_{1}$ & \multicolumn{2}{|c|}{1.33333}\\
 \hline
 &Weak  & Strong & & Weak	 & Strong  \\ [0.5ex] 
 \hline\hline
 $\alpha_2$	& 0.8942 & 1.40331 &$\alpha_2^\prime$	&0.35191  &   1.05019\\ 
 \hline
 $\beta_2$ & 0.737 & 0.9033 &$\beta_2^\prime$ & -0.73887 & -0.38151 \\
 \hline
 $\frac{p_2}{\rho_0c^2}$ & 7.127 & 18.8817 &$\frac{p_2}{\rho_0c^2}$ &  7.127 & 18.8817\\
 \hline
 $\rho_2/\rho_0$ & 16.3425 & 27.6949 &$\rho_2/\rho_0$ &  16.3425 & 27.6949  \\[1ex] 
 \hline
\end{tabular}
\caption{Weak and strong solutions of a shock collision.}
\label{table_weak_strong}
\end{table}

Figure~\ref{fig_S2Wtime_evolution} shows the resulting evolution of the mean density and pressure of the doubly-shocked region, in terms of its fractional deviation from the weak and strong shock solutions.  The system quickly transitions from the algebraic strong shock RR solution to its numerical counterpart, in which the density and pressure in region 2 differ by $\sim\!1.5\!-\!2\%$. The system then starts to linearly deviate from this solution with a growth rate of about $0.5~{t_{sc}^\prime}^{-1}$ or
$e$-folding time about $2t_{sc}^\prime$ (two sound crossing times). Subsequently, the transition between the strong and weak shock solutions enters a non-linear phase. Finally, 
the weak shock RR solution is approached at about 10 sound crossing time ($t_{sc}^\prime$), and the deviation from this solution appears to decrease exponentially with time. 

Figure~\ref{fig_S2Wsnaps_rhoprs} shows snapshots from the corresponding simulation (in frame S') in Figure~\ref{fig_S2Wsnaps_rhoprs} displaying the fluid variables at different times (indicated by the dashed vertical lines in Figure~\ref{fig_S2Wtime_evolution}). 
Panels b~\&~$\bar{b}$ show a small (linear order) change in density, pressure and shape of the doubly-shocked region~2 from its initial state (which is shown in panels a~\&~$\bar{\rm a}$). 
Panels c~\&~$\bar{c}$ indicate the significant (non-linear) changes with the transient appearance of a new (third) shock and a contact discontinuity, which bound a triply-shocked region at the bottom-right corner of the snapshot. 
Panels d~\&~$\tilde{d}$ show the system reached the weak shock solution and it stays there.

\subsubsection{Irregular Reflection (IR) or Mach Reflection (MR)}
In the sub-sonic region, if there was 
RR then the dense, high-pressure fluid in the doubly-shocked region 2 would be in causal contact with point P, and cause it to detach from the wall, thereby leading to IR. For this reason there is no RR in the sub-sonic region, and instead only IR types of shock reflection such as MR. As mentioned in \S\,\ref{sec:par_space}, there may be a dual region within the super-sonic region where both RR and IR/MR are possible, but the weak shock RR solution appears to be the most stable attractor solution that generically appears in our simulations.

Therefore, we generally expect the formation of IR/MR in our simulations in the sub-sonic regime. For a given $u_1$, this corresponds to sufficiently large incidence angles $\alpha_1$. 
For such incidence angles, the post-shock region develops multiple zones separated by discontinuities.

\begin{figure}
	\includegraphics[width=0.988
 \columnwidth]{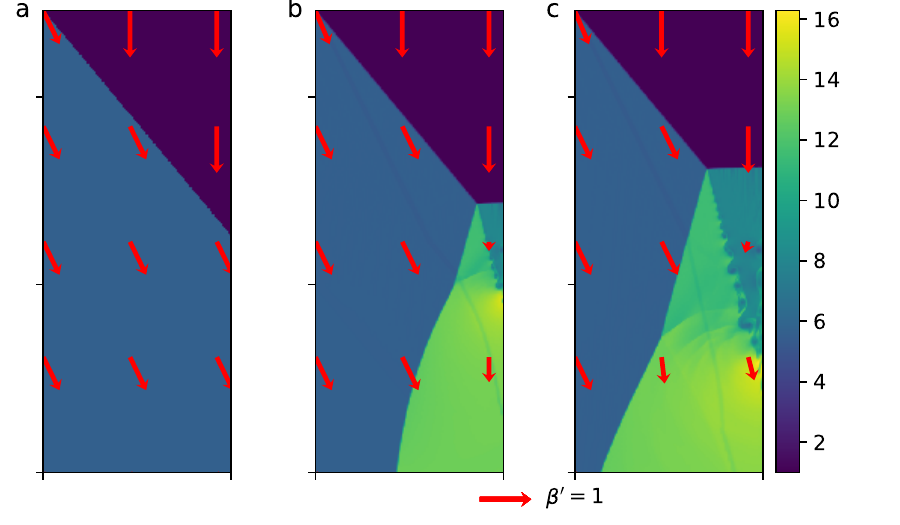}
    \caption{Sanpshots from a simulation with $(\alpha_1,\,u_1)=(1.1,\,1.0)$ in which Mach reflection (MR) develops, at times $t^\prime/t_{sc}^\prime=0$, $t^\prime/t_{sc}^\prime=0.2$, and $t^\prime/t_{sc}^\prime=0.4$.Each panel shows a colormap of the proper rest-mass density and red velocity vectors. 
    The side ratio of the computation domain is $L'_y/L'_x=2.46$. 
    }	
    \label{fig_ShockWall_Mov_mach}
\end{figure}

Figure~\ref{fig_ShockWall_Mov_mach} shows an example of a simulation for such a case, where MR develops.
For this numerical simulation we considered an incident shock characterized by $(\alpha_1,\,u_1)=(1.1,\,1.0)$, with an unshocked region~0 of ($\rho_0=1$, $p_0/\rho_0c^2=10^{-9}$, $u_0=0$). 
To calculate the fluid variables in frame S$^\prime$ we consider the corresponding boost of $\beta_p=0.9067722$, which implies an S' frame incidence angle of
$\tan\alpha_1^\prime=0.8283838$. We evolve the fluid maintaining the boundary conditions mentioned in section~\ref{sec_numerical_setup_Sprime}. 

Figure~\ref{fig_ShockWall_Mov_mach} shows snapshots from this simulation. 
The meeting point of the incident and the reflected shocks, P, detaches from the reflecting wall (\textit{panels} b, c) 
and a Mach stem develops behind which there is singly shocked fluid at pressure equilibrium across a contact discontinuity with doubly-shocked fluid behind the reflected shock s2. One can clearly see the development of Kelvin–Helmholtz instability along this contact discontinuity due to the velocity shear (discontinuous parallel velocity component) across it.
The reflected shock assumes
an irregular non-triangular shape.
The Mach stem slowly moves upward in frame S'  where this simulation is performed, at a constant speed, such that its length (or the distance of point P from the wall) increases linearly with time.
In this paper, we do not explore in detail the characteristics of this IR/MR and instead leave this for a future work.

\subsubsection{consistency of numerical results in frames S and S'}

\begin{figure}
	\includegraphics[width=0.988
 \columnwidth]{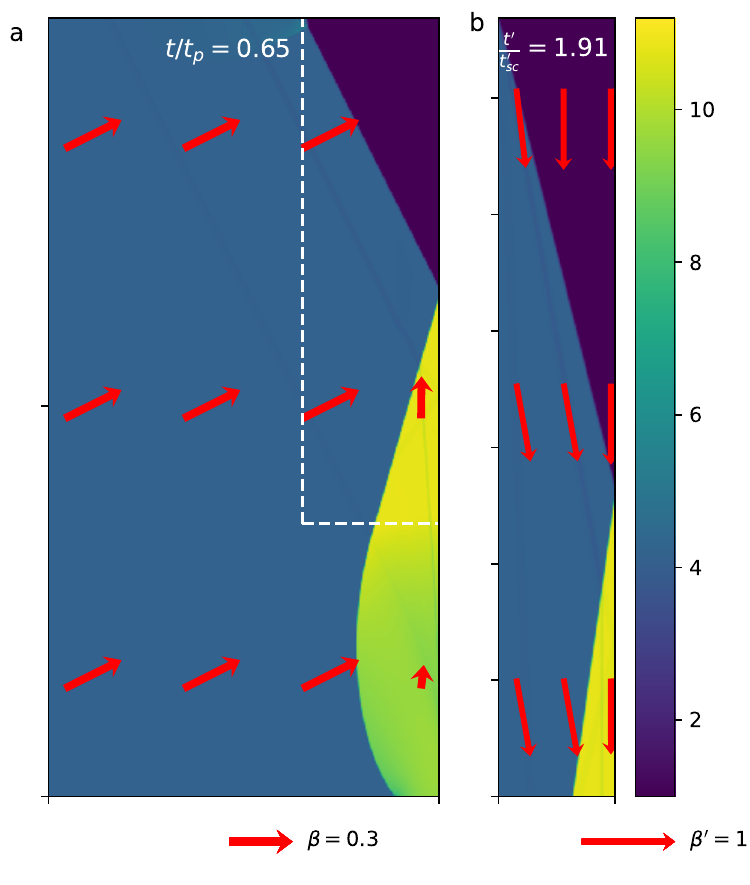}
    \caption{Snapshots (proper rest-mass density colormap and red velocity vectors) from simulations of RR for $(\alpha_1,\,u_1)=(0.465,\,0.316)$ performed in: (a) the lab frame S ($\tan\alpha_2=0.281$) at $t/t_p=0.65$, and (b) the moving frame S$^\prime$ ($\tan\alpha_1^\prime=0.246$, $\tan\alpha_2^\prime=0.139$) at $t'/t'_{sc}=1.91$}.	
    \label{fig_ShockWall_LabMov}
\end{figure}
Here we show the consistency of the numerical results obtained through relativistic hydrodynamic simulations performed in rest frames S and S$^\prime$. Figure~\ref{fig_ShockWall_LabMov} shows the snapshots from frames S and S$^\prime$ for the same physical shock reflection case of 
$(\alpha_1,\,u_1)=(0.465,\,0.316)$.
In the lab frame S the frame S$^\prime$ moves with velocity $v_p=0.87c$ upward along the wall, such that the incidence angle $\tan\alpha_1=0.501$ in S transforms to $\tan\alpha_1^\prime=0.246$ in S$^\prime$. From the numerical evolution studies, we obtain the development of the doubly-shocked region~2. In frame S we derive a value of $\tan\alpha_2=0.281$ for the angle of the reflected shock s2 relative to the wall, and in frame S$^\prime$ we derive a corresponding value of $\tan\alpha^\prime_2=0.139$. Using a Lorentz boost from frame S$^\prime$ to S we obtain $\tan\alpha_2=0.284$ from the evolution in frame S$^\prime$. The algebraic solution corresponding to the same input parameters gives $\tan\alpha_2=0.282355$. Therefore, the angle of reflection is consistent in both frames S and S$^\prime$ to within 1\%, and both are consistent with the analytic value.

\subsubsection{consistency of the numerical and analytic results}

\begin{figure}
\includegraphics[trim={0.20cm 1.2cm 1.2cm 2.1cm},clip,scale=0.7,width=0.48\textwidth]{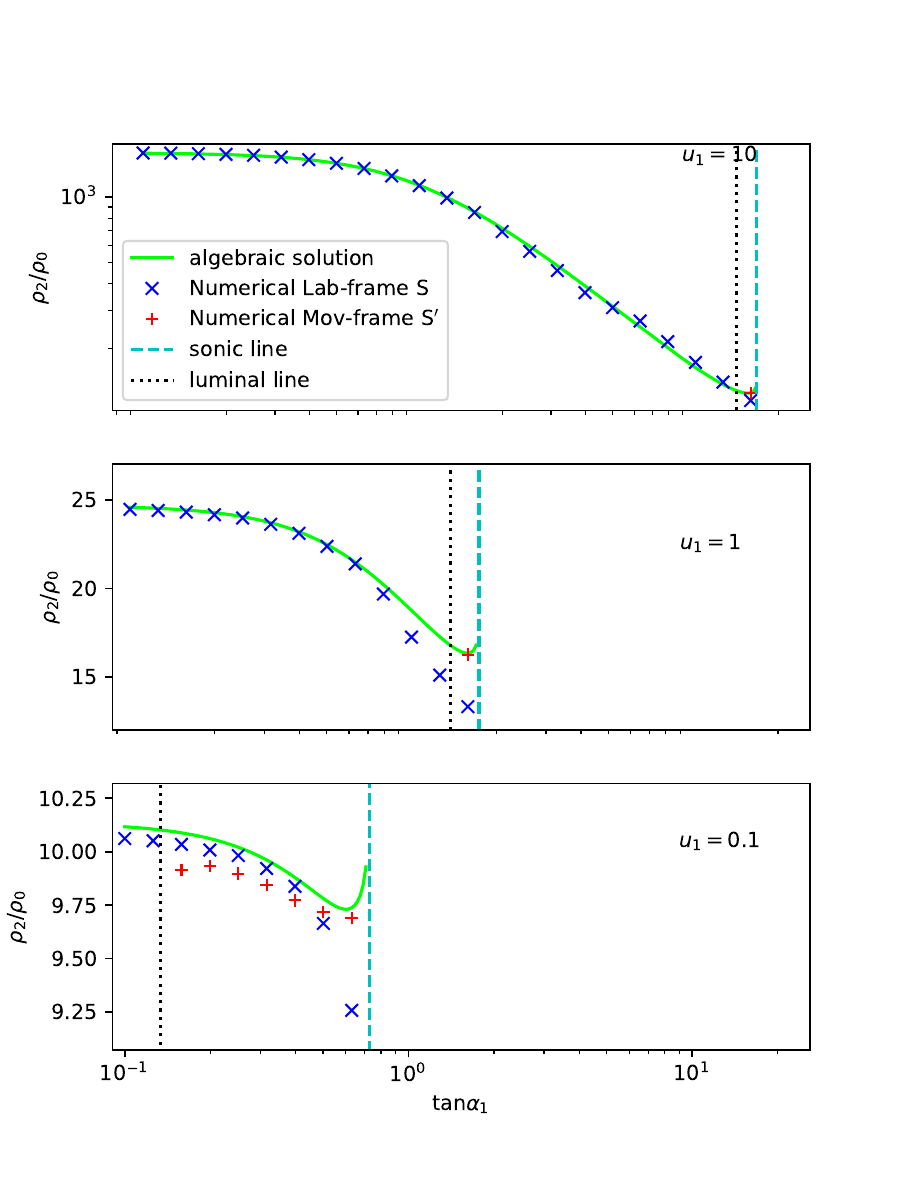}
    \caption{Comparing the matter proper rest-mass density in the doubly-shocked region~2 ($\rho_2$, normalized by $\rho_0$) 
    from the analytic algebraic solution (solid \green{green}  lines) to our hydrodynamic simulation results, in the lab frame S (blue \blue{x} symbols) and in the moving frame S$^\prime$ (red \red{+} symbols), for $u_1=10,1,0.1$ (from top to bottom). The vertical dotted black line (dashed \cyan{cyan} line) corresponds to the luminal (sonic) line. Frame S is well-suited for low 
    $\alpha_1$ values, while frame S$^\prime$ is a favorable option near the sonic line.
    }	
\label{fig_ShockWall_compare_rho2_regular}
\end{figure}
\begin{figure}
 \includegraphics[trim={0.25cm 1.2cm 1.2cm 2.1cm},clip,scale=0.7,width=0.48\textwidth]{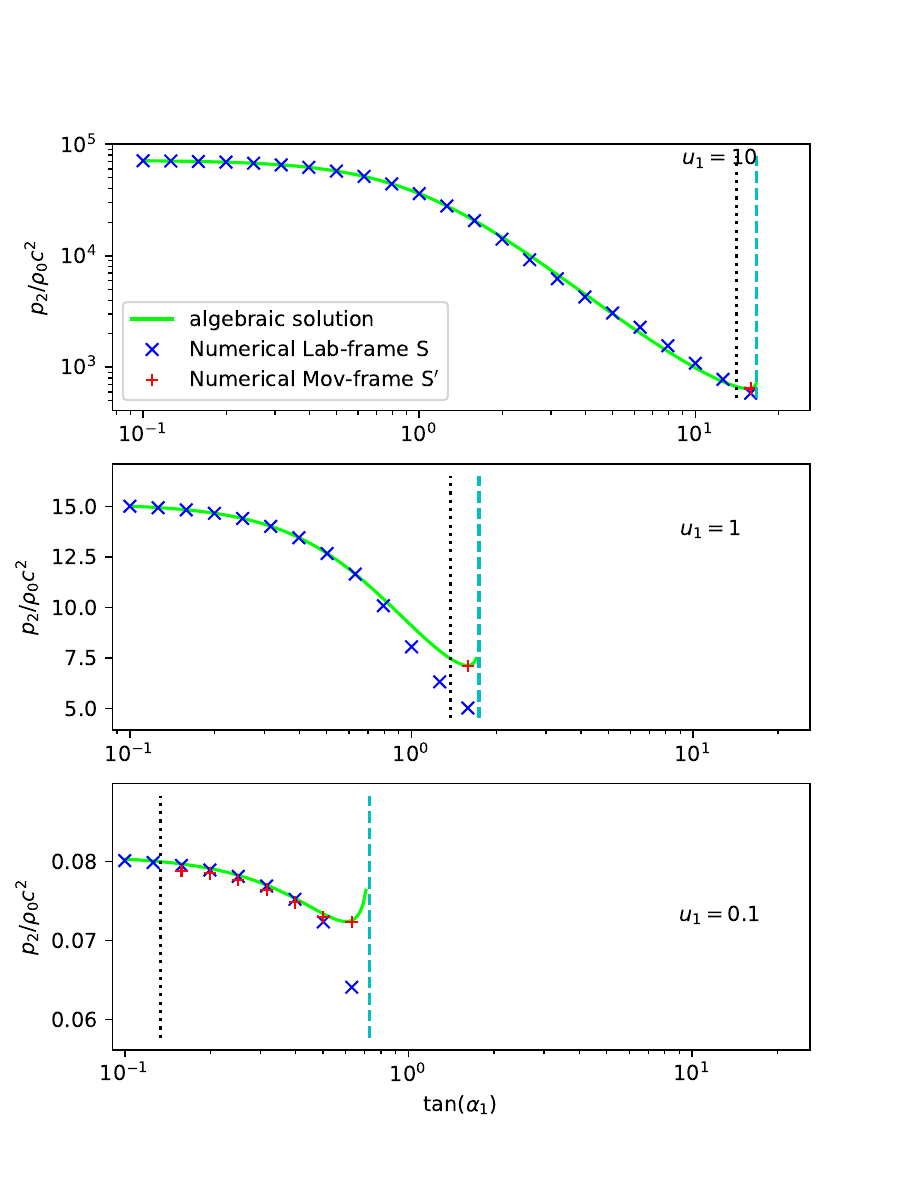}
    \caption{Similar to Fig.~\ref{fig_ShockWall_compare_rho2_regular} but comparing the region 2 pressure.
        }	
    \label{fig_ShockWall_compare_prs2_regular}
\end{figure}
\begin{figure}
	\includegraphics[trim={0.45cm 1.2cm 1.2cm 2.1cm},clip,scale=0.7,width=0.48\textwidth]{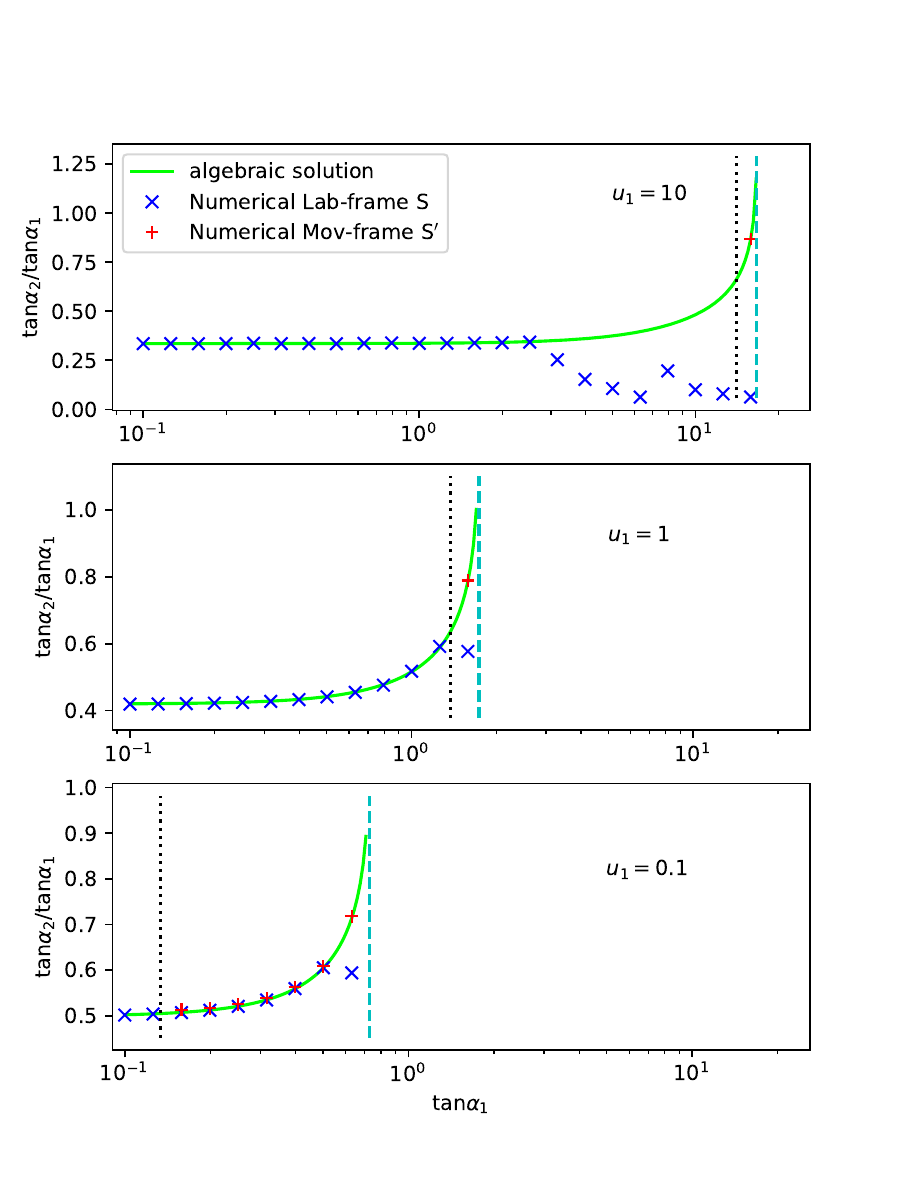}
    \caption{
    Similar to Fig.~\ref{fig_ShockWall_compare_rho2_regular} but comparing 
$\tan\alpha_2/\tan\alpha_1$. 
    }	
    \label{fig_ShockWall_compare_tanalpha2_regular}
\end{figure}

We compare the results of our numerical simulations with the exact algebraic solution for the relatively simpler case of RR, for which such an analytic solution can be obtained (GR23). 
The critical incidence angle along the sonic line, $\alpha_{1,{\rm sonic}}(u_1)$, below which RR is possible, increases as the incident shock velocity $\beta_{s1}$ increases. A detailed comparison is shown for the proper rest-mass density $\rho_2$ (Figure~\ref{fig_ShockWall_compare_rho2_regular}), and pressure $p_2$ (Figure~\ref{fig_ShockWall_compare_prs2_regular}) of the doubly-shocked region 2, as well as the ratio of the tangens of the angles relative to the wall of the reflected and incident shock fronts, $\tan\alpha_2/\tan\alpha_1$ (Figure~\ref{fig_ShockWall_compare_tanalpha2_regular}).

We note that lab frame S simulation is more accurate for a smaller incidence angle $\alpha_1$. As the value of $\alpha_1$ increases the doubly-shocked region~2 in frame S is more strongly affected by the  imposed inflow lower boundary condition. Hence, the numerical results deviate from the expected value as $\alpha_1$ approaches the critical value for a RR (i.e. the sonic line). Frame S$^\prime$ is more suitable for simulations in this regime, and can follow the shock reflection for longer times, as the flow becomes steady in S$^\prime$ for RR.

\subsection{The Sonic Line}

\subsubsection{The Significance of the Sonic Line}

The sonic line corresponds to the condition
\begin{equation}
\beta'_{2,w}=\beta_{c_s,2,w}\ \ \Longleftrightarrow \ \ \beta_p=\frac{\beta_{2,w}+\beta_{c_s,2,w}}{1+\beta_{2,w}\beta_{c_s,2,w}}\ .
\end{equation}
The first condition is that in the rest frame S$^\prime$ where the flow is steady the velocity in region 2 (of the doubly-shocked fluid) for the weak shock RR solution, $\beta'_{2,w}$, is equal to its sound speed, $\beta_{c_s,2,w}$. Once $\beta'_{2,w}$ drops below $\beta_{c_s,2,w}$, i.e. in the subsonic regime,  region 2 comes into causal contact with point $P$, and can then potentially cause it to separate from the wall resulting in MR. On the other hand, in the super-sonic regime ($\beta'_2>\beta_{c_s,2}$) region 2 is not in causal contact with point $P$ (for an initial unperturbed weak shock RR solution) so it cannot affect it and therefore point $P$ cannot separate from the wall and allow a transition to IR/MR. This may potentially suppress a transition between the weak shock RR solution and MR (in the dual region between the sonic line and the mechanical equilibrium line; see e.g. GR23), and require a sufficiently large perturbation for it to occur. The fact that the strong shock RR solution is always subsonic ($\beta'_{2,s}<\beta_{c_s,2,s}$)
may potentially account for its instability, e.g. as found in \S\,\ref{subsec_regular_reflection}.
Since the sonic condition is that for causality between region 2 and point $P$, it can also be expressed in the lab frame S such that the speed of a sound wave propagating in region 2 along the wall towards point $P$, $(\beta_2+\beta_{c_s,2})/(1+\beta_2\beta_{c_s,2})$, equals that of point $P$, $\beta_p$. 

\subsubsection{The Flow Properties along the Sonic Line}

The sonic line's physical significance makes it interesting to study in detail the flow properties along it.
The analytic solution for the flow properties along the sonic line is derived in Appendix~\ref{sec:Appendix-sonic-line}, along with analytic expressions for all of the flow quantities in the Newtonian and relativistic limits.

\begin{figure}
	\includegraphics[width=1.0\columnwidth]{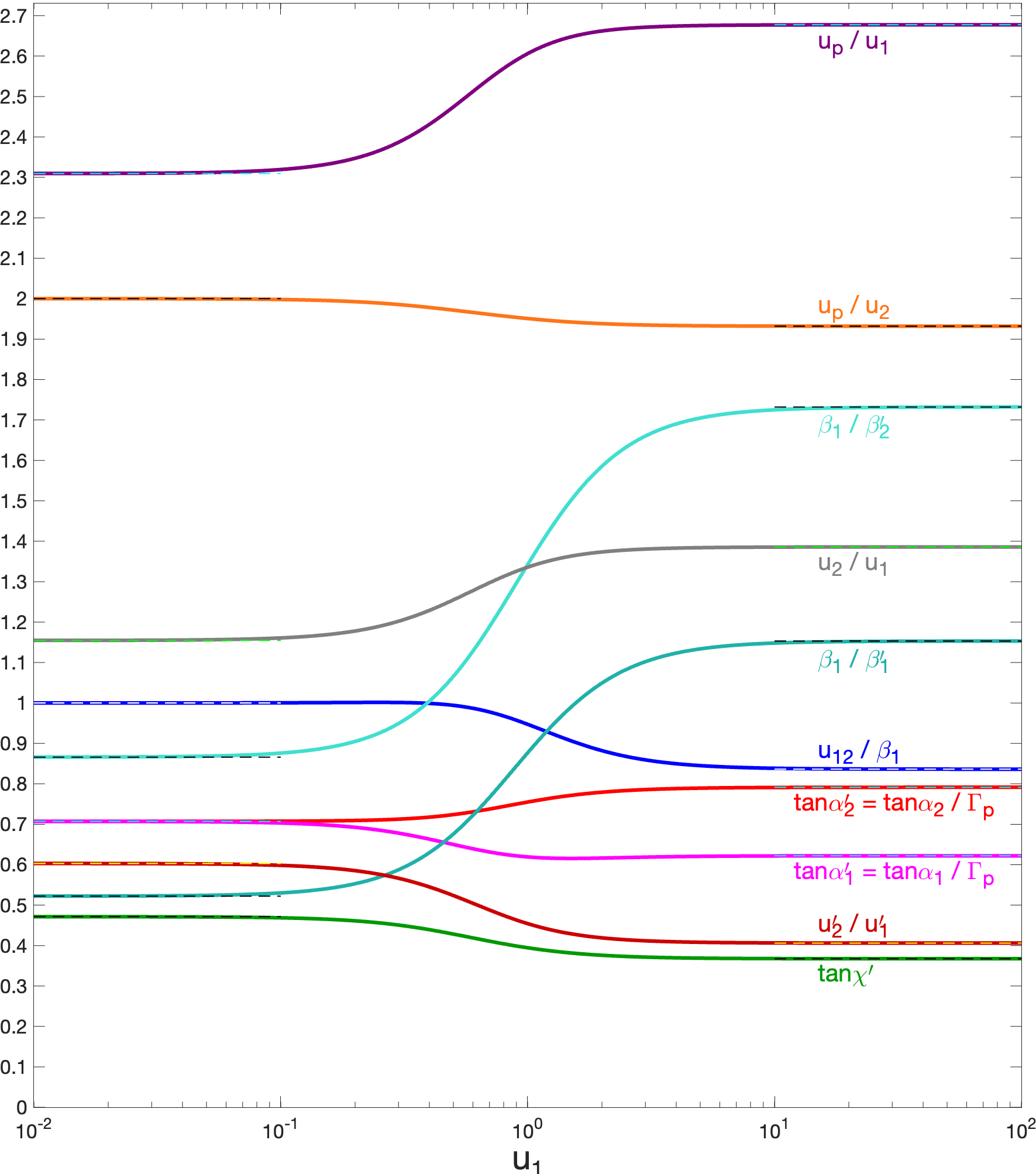}
    \caption{The values of different hydrodynamic variables along the sonic line ($\beta'_2=\beta_{c_s,2}$) are shown as a function of the value of $u_1$ along this line (\textit{thick solid lines}). The Newtonian and relativistic limits from Equations~(\ref{eq:sonic_Newt}) and (\ref{eq:sonic_rel}), respectively, are indicated by thin dashed lines (which fall on top of the thick solid lines).}	
    \label{fig:sonic_line_an}
\end{figure}

Figure~\ref{fig:sonic_line_an} shows the values of different hydrodynamic variables along the sonic line ($\beta'_{2,w}=\beta_{c_s,2,w}$), conveniently parameterised according to the value of $u_1$ along this line. These values are found by numerically solving the set of algebraic equations for RR, together with the sonic condition, in the frame S$^\prime$ (Appendix~\ref{sec:Appendix-sonic-line}) or equivalently in the lab frame S (as is done in GR23). The weak and strong shock RR solutions exactly coincide at the detachment line, which almost coincides with the sonic line, such that both solutions are extremely close along the sonic line. There is excellent agreement with both the Newtonian and the relativistic limits that are found analytically in Appendix~\ref{sec:Appendix-sonic-line}. 
Moreover, it can be seen from Fig.~\ref{fig:sonic_line_an} that in the relativistic limit $u_p>u_2>u_1\gg1$ while $u_{12}$ is of order unity, such that the first (incident) shock $s1$ is ultra-relativistic, while the second (reflected) shock $s2$ is mildly relativistic.

\begin{figure}
	\includegraphics[width=0.97\columnwidth]{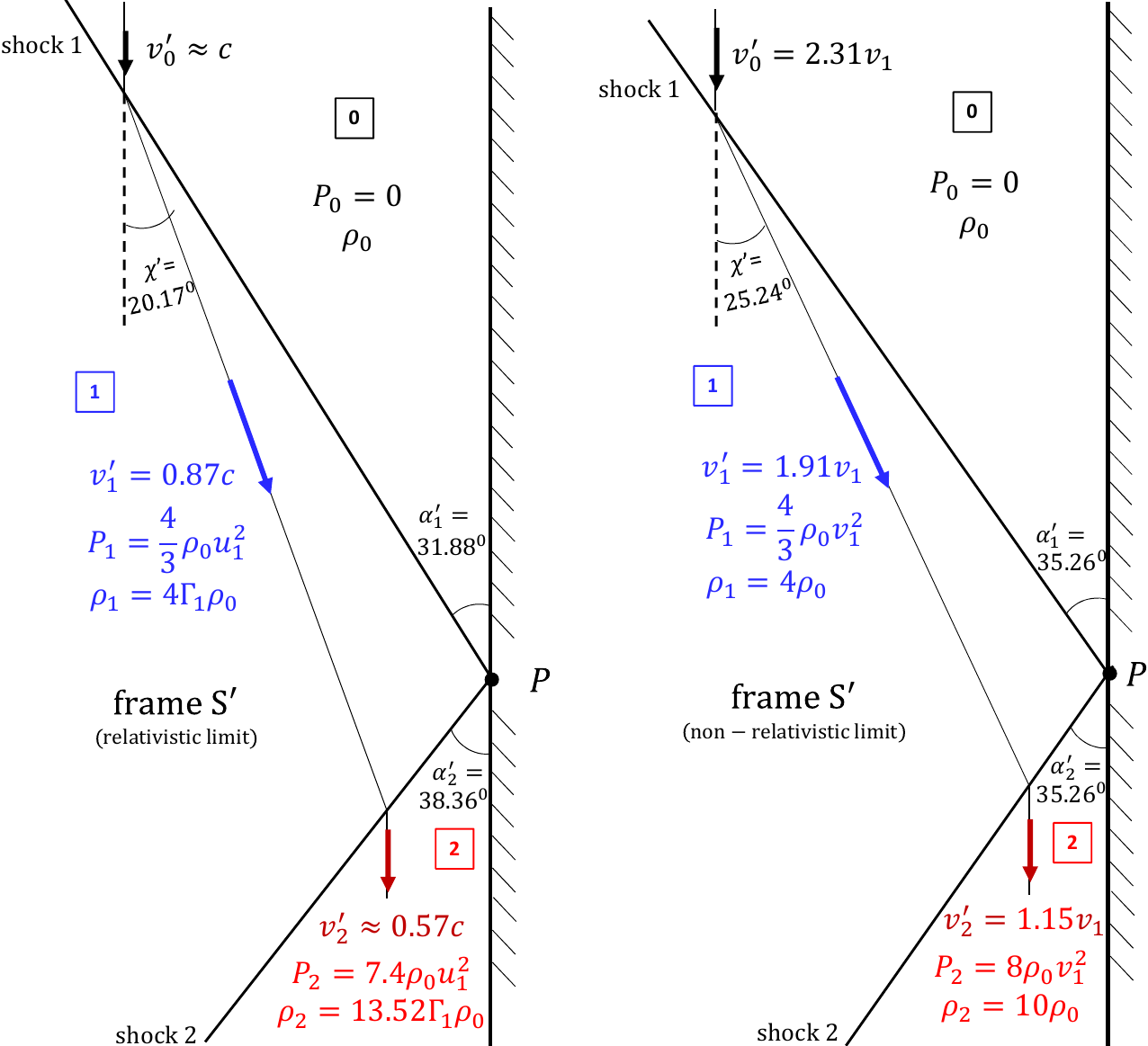}
    \caption{The asymptotic Newtonian (\textit{right panel}) and relativistic (\textit{left panel}) flow configurations along the sonic lines, for which the flow parameters are given in Eqs.~(\ref{eq:sonic_Newt}) and (\ref{eq:sonic_rel}), respectively.}	
    \label{fig:sonic_line_limits}
\end{figure}

\begin{figure}
	\includegraphics[width=0.97\columnwidth]{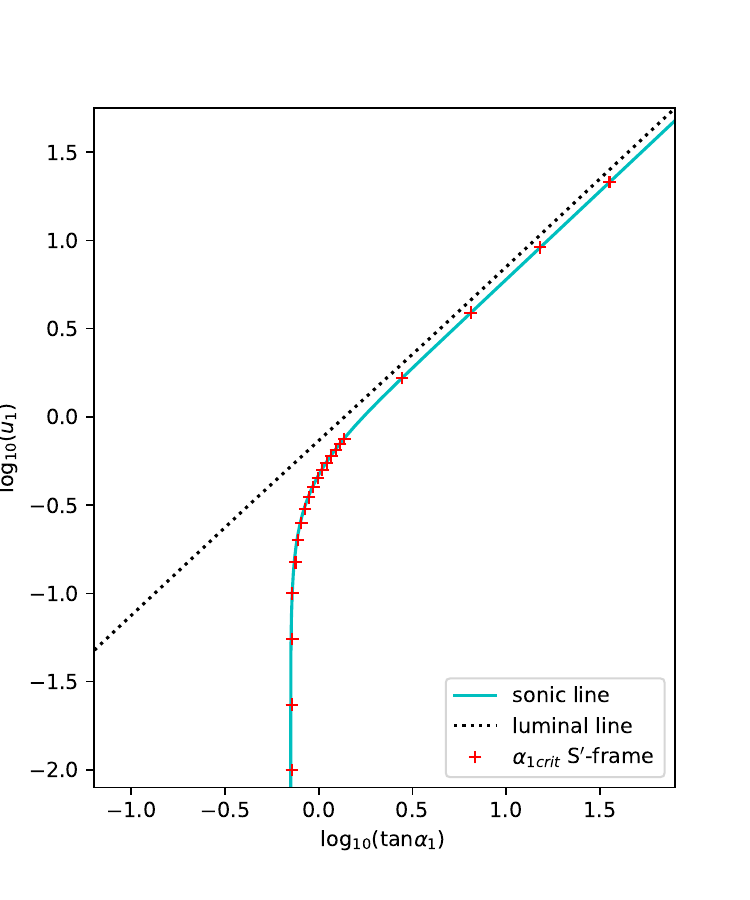}
    \caption{The location of the transition from RR to MR (red \red{+} symbols; $\alpha_{1,{\rm crit}}(u_1)$) found from our numerical simulations in frame S$^\prime$, match the analytically calculated location of the sonic line (solid \cyan{cyan} line; $\alpha_{1,{\rm sonic}}(u_1)$). The dotted black line indicates the luminal line, which is shown for reference.
    }	
    \label{fig:sonic_line}
\end{figure}

\begin{figure}
	\includegraphics[width=0.97\columnwidth]{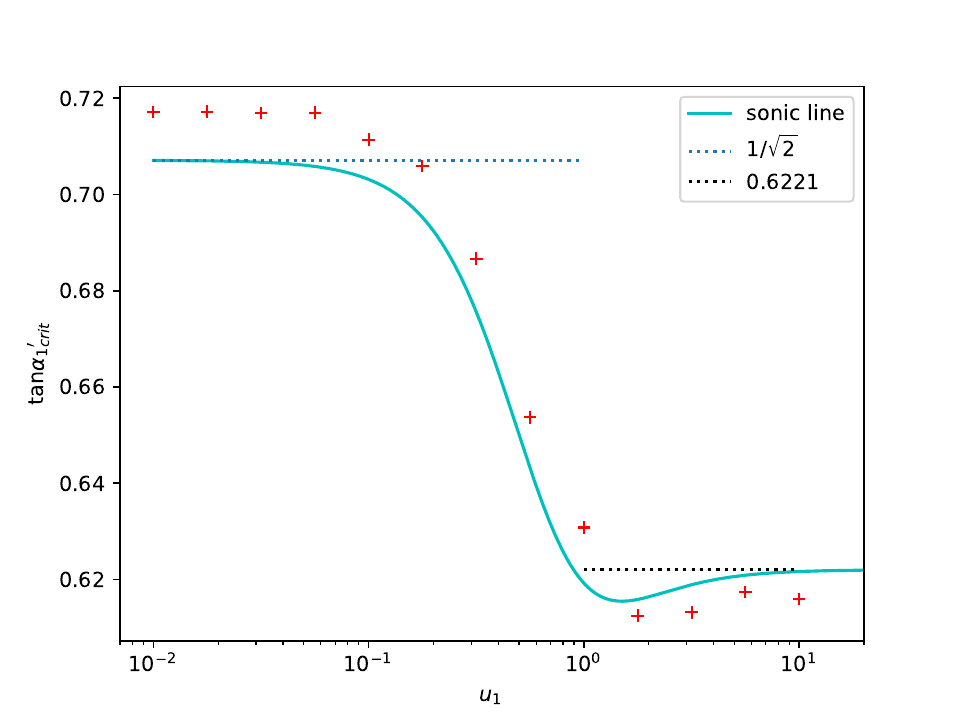}
    \caption{The critical transition angle in frame S$^\prime$, $\tan\alpha'_{1,{\rm crit}}(u_1)$, from our numerical simulations (red \red{+} symbols), compared to the analytically derived corresponding angle for the sonic line, $\tan\alpha'_{1,{\rm sonic}}(u_1)$, whose asymptotic Newtonian and relativistic limits are shown by horizontal black dotted lines.  
    }	
    \label{fig_sonic_line_u_range}
\end{figure}

Figure~\ref{fig:sonic_line_limits} shows the flow configurations for the asymptotic Newtonian and relativistic limits along the sonic line, in the rest frame S$^\prime$ where the flow is steady and point $P$ is at rest. These limits are particularly simple and can be fully solved analytically (Appendix~\ref{sec:Appendix-sonic-line}). In frame S$^\prime$ the angles $\alpha'_1$, $\alpha'_2$ and $\chi'$ do not vary that drastically between these two limits (see also Fig.~\ref{fig:sonic_line_an}).

Figure~\ref{fig:sonic_line} shows the location of the transition from RR to MR (red \red{+} symbols). For each $u_1$ values the incidence angle $\alpha_1$ is gradually increased between different simulations until we find the critical value $\alpha_{1,{\rm crit}}(u_1)$ at which point P detaches from the wall, signaling the transition from RR to MR. It is determined more accurately by performing several iterations for each $u_1$ value. These numerical values are in excellent agreement with the analytically calculated location of the sonic line (solid \cyan{cyan} line in Fig.~\ref{fig:sonic_line}), $\alpha_{1,{\rm crit}}(u_1)=\alpha_{1,{\rm sonic}}(u_1)$. These simulations clearly shows that: (i) there is indeed no RR in the sub-sonic region, and (ii) in the super-sonic region the simulations reach the weak shock RR solution and not the strong shock RR solution (in the sub-luminal region) or MR (in the dual region).

Figure~\ref{fig_sonic_line_u_range} shows the critical transition angle in frame S$^\prime$, $\tan\alpha'_{1,{\rm crit}}(u_1)$, from our numerical simulations (red \red{+} symbols), compared to the analytically derived corresponding angle for the sonic line, $\tan\alpha'_{1,{\rm sonic}}(u_1)$, whose asymptotic Newtonian and relativistic limits are shown by horizontal black dotted lines. They agree to within about $1\%$. This critical angle in frame S$^\prime$ decreases by about $10\%$ between the Newtonian and relativistic limits (see also Fig.~\ref{fig:sonic_line_an}).

\section{Conclusions}\label{sec:conclusions}

We have studied relativistic shock reflection, mainly numerically using two dimensional relativistic hydrodynamic simulations, and with detailed comparisons to analytic results. Our simulations were performed in two different rest frames: the lab frame S where the cold unshocked fluid (region~0) is at rest, and the rest frame S$^\prime$ where for RR the point P of intersection of the incident and reflected shocks with the reflecting wall is at rest, and the flow is in a steady state.

Our numerical simulations have validated the analytic results (derived mainly in GR23, but also in Appendix~\ref{sec:Appendix-sonic-line} for the sonic line). We have also pointed out the importance of using a suitable reference frame in different simulations of shock interactions. In particular, for small incidence angles $\alpha_1$
the lab frame S
is more suitable, while the frame S$^\prime$ that exists only in the sub-luminal region is more suitable closer to the sonic line.

We have also studied the transition between RR and IR (namely MR). 
The transition from RR to IR/MR maintains similar characteristics in the Newtonian and relativistic regimes. In our numerical study, this transition occurred exactly at the sonic/detachment line.

Moreover, in the super-sonic region the simulations always reached the weak shock RR solution. We have found the alternative strong shock RR solution, which exists in the super-sonic sub-luminal region, to be unstable. Moreover, we numerically studied how it transitions to the weak shock RR solution, which appears to be a stable attractor solution. While a dual region where either RR or MR are possible should exist from analytic considerations (on the super-sonic side of the sonic line but well within the sub-luminal region), it was never reached in our simulations, suggesting that it is not an attractor solution (and may also be unstable).

\section*{acknowledgement}
P. Bera is supported by the Israel Academy of Sciences and Humanities \& Council for Higher
Education Excellence Fellowship Program for International Postdoctoral Researchers.
This research was funded in part by the ISF-NSFC joint research program under grant no. 3296/19 (J.G.) and by the United States-Israel Binational Science Foundation (BSF) under grant no. 2020747 (P. Beniamini).

\section*{Data availability}
The data underlying this article will be shared on reasonable request to the corresponding author.
 
\appendix 
\section{The Solution along the Sonic Line}
\label{sec:Appendix-sonic-line}

Here we derive an analytic solution for RR along the sonic line. 
The conditions in region 1, for a cold region 0 ($p_0=0$, $e_0=w_0=\rho_0c^2$ and $h_0=1$), can be conveniently calculated in the lab frame S and are given by
\begin{eqnarray}\nonumber
&\rho_1=4\Gamma_1\rho_0\ ,\quad
p_1 = \frac{4}{3}u_1^2\rho_0c^2\ ,\quad
e_1=4\Gamma_1^2\rho_0c^2\ ,\quad
\\
&e_{\rm int,1} = \frac{4\Gamma_1u_1^2}{\Gamma_1+1}\rho_0c^2\ ,\quad 
w_1 = 4\Gamma_1^2\left(1+\frac{\beta_1^2}{3}\right)\rho_0c^2\ ,
\\ \nonumber
&\beta_{s1} = \frac{4\Gamma_1u_1}{4\Gamma_1^2-1}\ ,\quad
u_1 = \frac{1}{2}\sqrt{u_{s1}^2-2+\sqrt{4+5u_{s1}^2+u_{s1}^4}}\ ,
\\ \nonumber
&\beta_{1,s1} = \frac{\beta_{s1}-\beta_1}{1-\beta_1\beta_{s1}} = \frac{\beta_1}{3}\ ,
\end{eqnarray}
(GR23) where the last equation means that for our equation of state, in the rest frame of the downstream fluid (region 1), the speed at which the shock is receding is a third of the incoming upstream speed. 

Since the sonic line is always in the sub-luminal regime, it can conveniently be analyzed in frame S$^\prime$. In this frame the flow is steady and there are two oblique shocks: $s1$ and $s2$, at angles $\alpha'_1$ and $\alpha'_2$, respectively, relative to the wall. 
The velocity of region 1 in frame S$^\prime$ can be expressed through
\begin{eqnarray}\nonumber
\textbf{\textit{u}}'_1&=&\left[-u_1\cos\alpha_1,\;\Gamma_p\Gamma_1(\beta_1\sin\alpha_1-\beta_p)\right]\ ,\quad\quad
\\ \nonumber
\Gamma'_1 &=& \sqrt{1+u^{\prime\,2}_1} = \Gamma_1\Gamma_p(1-\beta_1\beta_p\sin\alpha_1)
\\ 
 &=&\frac{3\,\Gamma_1\sin\alpha_1}{\sqrt{(4\Gamma_1^2-1)^2\sin^2\alpha_1-16\Gamma_1^2u_1^2}}\ ,\quad\quad
\\ \nonumber
\tan\chi' &=& \frac{u'_{1x}}{u'_{1y}} = \frac{\beta_1\cos\alpha_1}{\Gamma_p(\beta_p-\beta_1\sin\alpha_1)}\ ,
\\ \nonumber
\tan\alpha'_1 &=&\frac{\tan\alpha_1}{\Gamma_p}\ ,\quad\quad\quad\tan\alpha'_2 =\frac{\tan\alpha_2}{\Gamma_p}\ .
\end{eqnarray}
The remaining conditions are the oblique shock jump conditions in frame S$^\prime$ and the sonic condition, which read
\begin{eqnarray}\nonumber
\rho_1u'_1\sin\alpha'_+&=&\rho_2u'_2\sin\alpha'_2\ ,
\\ \nonumber
w_1u^{\prime\,2}_1\sin^2\alpha'_++p_1&=&w_2u^{\prime\,2}_2\sin^2\alpha'_2+p_2\ ,
\\ \label{eq:Sp0}
w_1\Gamma'_1u'_1\sin\alpha'_+&=&w_2\Gamma'_2u'_2\sin\alpha'_2\ ,
\\ \nonumber
\beta'_1\cos\alpha'_+&=&\beta'_2\cos\alpha'_2\ ,
\\ \nonumber
\beta'_2 &=& \beta_{c_s,2}\ ,
\end{eqnarray}
where we denote $\alpha'_+=\chi'+\alpha'_2$ and
\begin{equation}
\beta_{c_s,2}^2=\frac{3\Theta_2^2+5\Theta_2\sqrt{\Theta_2^2+4/9}}{12\Theta_2^2+2+12\Theta_2\sqrt{\Theta_2^2+4/9}}\ ,\quad\quad\Theta_2=\frac{p_2}{\rho_2c^2}\ .
\end{equation}

In the Newtonian limit ($\beta_1<\beta_p\ll1$) the adiabatic index is $\hat{\gamma}=5/3$ and the equations reduce to
\begin{eqnarray}\nonumber
&\frac{\rho_1}{\rho_0}=4\ ,\quad
\frac{p_1}{\rho_0} = \frac{4}{3}v_1^2\ ,\quad 
\frac{e_{\rm int,1}}{\rho_0} = 2v_1^2\ ,\quad
\beta_{s1}=\frac{4}{3}\beta_1 \ ,
\\ \nonumber
&\textbf{\textit{v}}'_1=v_1\left(-\cos\alpha_1,\;\;\sin\alpha_1-\frac{4}{3\sin\alpha_1}\right)\ ,%
\\
&\tan\chi' = \frac{v'_{1x}}{v'_{1y}} = \frac{3\cos\alpha_1\sin\alpha_1}{1+3\cos^2\alpha_1}\ ,
\\ \nonumber
\nonumber
&\rho_1v'_1\sin\alpha'_+=\rho_2v'_2\sin\alpha'_2\ ,
\\ \nonumber
&\rho_1v^{\prime\,2}_1\sin^2\alpha'_++p_1=\rho_2v^{\prime\,2}_2\sin^2\alpha'_2+p_2\ ,
\\ \nonumber
&5\frac{p_1}{\rho_1}+v^{\prime\,2}_1\sin^2\alpha'_+=5\frac{p_2}{\rho_2}+v^{\prime\,2}_2\sin^2\alpha'_2\ ,
\\ \nonumber
&v'_1\cos\alpha'_+=v'_2\cos\alpha'_2\ ,
\\ \nonumber
&v^{\prime\,2}_2 = \frac{5}{3}\frac{p_2}{\rho_2}\ ,
\end{eqnarray}
which have the following simple solution:
\begin{eqnarray}\nonumber
& v_p = v'_0 = \frac{4}{\sqrt{3}}v_1 
=\sqrt{3}\,v_{s1} = \frac{4}{\sqrt{11}}v'_1 = 2v'_2=2v_2\ ,
\\ \nonumber
& \rho_2 = \frac{5}{2}\rho_1 = 10\rho_0\ ,\quad\quad
p_2=6p_1=8\rho_0v_1^2\ ,
\\ \nonumber
&\frac{u_2}{u_1} \to \frac{\beta_2}{\beta_1} =\frac{2}{\sqrt{3}}\ ,\quad\quad
\frac{u'_2}{u'_1}\to\frac{\beta'_2}{\beta'_1} = \frac{2}{\sqrt{11}}\ ,
\\ \label{eq:sonic_Newt} 
&\frac{u_{12}}{u_1} \to  \frac{\beta_{12}}{\beta_1} = 1\ ,\quad\quad
\frac{u_p}{u_1} \to \frac{\beta_p}{\beta_1}= \frac{4}{\sqrt{3}}\ ,
\\  \nonumber
&\tan\alpha'_1 = \tan\alpha_1 = \tan\alpha'_2=\tan\alpha_2= \frac{1}{\sqrt{2}}\ ,
\\  \nonumber
&\sin\alpha'_1 = \sin\alpha_1 = \sin\alpha'_2=\sin\alpha_2=\frac{1}{\sqrt{3}}\ ,
\\  \nonumber
&\cos\alpha'_1 = \cos\alpha_1 =\cos\alpha'_2=\cos\alpha_2=\sqrt{\frac{2}{3}}\ ,
\\  \nonumber
&\tan\chi' = \frac{\sqrt{2}}{3}\ ,\quad\quad
\sin\chi' = \sqrt{\frac{2}{11}}\ ,\quad\quad
\cos\chi' = \frac{3}{\sqrt{11}}\ ,
\end{eqnarray}

From Fig.~\ref{fig:sonic_line_an} it can be seen that in the relativistic limit ($u_p>u_2>u_1\gg1$)  $u_{12}$ is of order unity, such that while the first (incident) shock $s1$ is ultra-relativistic (with a relative upstream to downstream proper speed of $u_1\gg1$), the second (reflected) shock $s2$ is only mildly relativistic.
Therefore, while region 0 is cold, both regions 1 and 2 are relativistically hot, with an adiabatic index of $\hat{\gamma}=4/3$ and $p=e_{\rm int}/3\gg\rho c^2$ while $w\approx 4p$ and $\beta_{c_s,2}=1/\sqrt{3}$, such that the sonic condition implies $\beta'_2=1/\sqrt{3}$, $u'_2=1/\sqrt{2}$, $\Gamma'_2=\sqrt{3/2}$. This also implies $\frac{p_2}{p_1}\approx\frac{e_{2,{\rm int}}}{e_{1,{\rm int}}}\approx\frac{ e_2}{e_1}\approx\frac{w_2}{w_1}$. Therefore, in the relativistic limit along the sonic line the equations reduce to:
\begin{eqnarray}\nonumber
&\frac{\rho_1}{\rho_0}=4\Gamma_1\ ,\quad \frac{4w}{3\rho_0c^2}\approx\frac{e_{1,{\rm int}}}{\rho_0c^2}\approx\frac{e_1}{\rho_0c^2}=4\Gamma_1^2\ ,
\\ \nonumber
&\frac{p_1}{\rho_0c^2}=\frac{4}{3}u_1^2\approx\frac{4}{3}\Gamma_1^2\ ,\quad
u_{s1} \approx  \sqrt{2}u_1\ ,\quad
\beta_{1,s1} = \frac{\beta_1}{3}\approx\frac{1}{3}\ ,
\\ \nonumber
&\rho_1u'_1\sin\alpha'_+=\rho_2\frac{1}{\sqrt{2}}\sin\alpha'_2\ ,
\\ \nonumber
&4p_1u^{\prime\,2}_1\sin^2\alpha'_++p_1=2p_2\sin^2\alpha'_2+p_2\ ,
\\ \label{eq:Sp0}
&4p_1\Gamma'_1u'_1\sin\alpha'_+=2\sqrt{3}p_2\sin\alpha'_2\ ,
\\ \nonumber
&\beta'_1\cos\alpha'_+=\frac{1}{\sqrt{3}}\cos\alpha'_2\ .
\end{eqnarray}
In the relativistic limit $\alpha_1,\,\alpha_2\approx\frac{\pi}{2}$ along the sonic line, so it is convenient to use the angle $\bar{\alpha}_1=\frac{\pi}{2}-\alpha_1$ 
to express the solution to the above equations in terms of 
\begin{eqnarray}\nonumber
a\equiv\Gamma_1\bar{\alpha}_1&\to&\sqrt{\frac{49\sqrt{3}+\sqrt{6195-3576\sqrt{3}-84}}{16(9-5\sqrt{3})}}
\\
&\approx & 0.6004187327198\ ,
\end{eqnarray}
where $a=\Gamma_1\bar{\alpha}_1\approx u_1\bar{\alpha}_1\approx u_1\cos\alpha_1\approx u_1/\tan\alpha_1$ approaches a constant values in this relativistic limit,
\begin{eqnarray}\nonumber
\frac{\Gamma_p}{\Gamma_1}&\to&\frac{u_p}{u_1}\to\sqrt{\frac{2}{1-2a^2}}\approx2.677423238004\ ,
\\ \nonumber
\frac{\Gamma_p}{\Gamma_2}&\to&\frac{u_p}{u_2}\to\frac{\sqrt{2}}{\sqrt{3}-1}\approx1.931851652578\ ,
\\ \nonumber
\frac{\Gamma_2}{\Gamma_1}&\to&\frac{u_2}{u_1}\to\frac{\sqrt{3}-1}{\sqrt{1-2a^2}}\approx1.385936251591\ ,
\\ \nonumber
\Gamma_{12}&\to&\frac{3\sqrt{3}-1-4a^2}{4\sqrt{1-2a^2}}\approx1.303551928752944\ ,
\\ \nonumber
\beta'_1&\to&\frac{1}{3}\sqrt{1+16a^2}\approx0.867182056601\ ,
\\ \nonumber
u'_1&\to&\sqrt{\frac{1+16a^2}{8(1-2a^2)}}\approx1.741360042442\ ,
\\ \nonumber
\Gamma'_1&\to&3/\sqrt{8(1-2a^2)}\approx2.008067428503\ ,
\\ \nonumber
\sin\alpha'_1 &\to& \sqrt{1-2a^2}\approx0.528199480119\ ,
\\ \nonumber
\cos\alpha'_1 &\to&\sqrt{2}a\approx0.849120314915\ ,
\\ \label{eq:sonic_rel}
\tan\alpha'_1 &\to& \sqrt{1-2a^2}/\sqrt{2}a\approx0.622054932430\ ,
\\ \nonumber
\sin\alpha'_2 &\to&\frac{1-\sqrt{3}+4a^2}{\sqrt{2(2-\sqrt{3})(1+4a^2)}}\approx0.620609996819\ ,
\\ \nonumber
\cos\alpha'_2 &\to&\frac{a\sqrt{8(1-2a^2)}}{\sqrt{2(2-\sqrt{3})(1+4a^2)}}\approx0.784119398975\ ,
\\ \nonumber
\tan\alpha'_2 &\to&\frac{1-\sqrt{3}+4a^2}{a\sqrt{8(1-2a^2)}}\approx0.791473846497\ ,
\\ \nonumber\sin\chi' &\to&\frac{a\sqrt{8(1-2a^2)}}{\sqrt{1+16a^2}}\approx 0.344798730926\ ,
\\ \nonumber\cos\chi' &\to&\frac{1+4a^2}{\sqrt{1+16a^2}}\approx 0.938676640357\ ,
\\ \nonumber\tan\chi' &\to&\frac{a\sqrt{8(1-2a^2)}}{1+4a^2}\approx 0.367324290498\ ,
\\ \nonumber
\frac{\rho_2}{\rho_1}&\to& \frac{1-\sqrt{3}+4(4-\sqrt{3})a^2}{2(1-\sqrt{3}+4a^2)\sqrt{1-2a^2}}\approx3.384471042815\ ,
\\ \nonumber
\frac{p_2}{p_1}&\to& \frac{\sqrt{3}-3+4(4\sqrt{3}-3)a^2}{4(1-\sqrt{3}+4a^2)(1-2a^2)}\approx5.549111674227\ .
\end{eqnarray}


\def\aj{AJ}%
\def\actaa{Acta Astron.}%
\def\araa{ARA\&A}%
\def\apj{ApJ}%
\def\apjl{ApJ}%
\def\apjs{ApJS}%
\def\ao{Appl.~Opt.}%
\def\apss{Ap\&SS}%
\def\aap{A\&A}%
\def\aapr{A\&A~Rev.}%
\def\aaps{A\&AS}%
\def\azh{AZh}%
\def\baas{BAAS}%
\def\bac{Bull. astr. Inst. Czechosl.}%
\def\caa{Chinese Astron. Astrophys.}%
\def\cjaa{Chinese J. Astron. Astrophys.}%
\def\icarus{Icarus}%
\def\jcap{J. Cosmology Astropart. Phys.}%
\def\jrasc{JRASC}%
\def\mnras{MNRAS}%
\def\memras{MmRAS}%
\def\na{New A}%
\def\nar{New A Rev.}%
\def\pasa{PASA}%
\def\pra{Phys.~Rev.~A}%
\def\prb{Phys.~Rev.~B}%
\def\prc{Phys.~Rev.~C}%
\def\prd{Phys.~Rev.~D}%
\def\pre{Phys.~Rev.~E}%
\def\prl{Phys.~Rev.~Lett.}%
\def\pasp{PASP}%
\def\pasj{PASJ}%
\def\qjras{QJRAS}
\def\rmxaa{Rev. Mexicana Astron. Astrofis.}%
\def\skytel{S\&T}%
\def\solphys{Sol.~Phys.}%
\def\sovast{Soviet~Ast.}%
\def\siamr{SIAMR}%
\def\ssr{Space~Sci.~Rev.}%
\def\zap{ZAp}%
\def\nat{Nature}%
\def\iaucirc{IAU~Circ.}%
\def\aplett{Astrophys.~Lett.}%
\def\apspr{Astrophys.~Space~Phys.~Res.}%
\def\bain{Bull.~Astron.~Inst.~Netherlands}%
\def\fcp{Fund.~Cosmic~Phys.}%
\def\gca{Geochim.~Cosmochim.~Acta}%
\def\grl{Geophys.~Res.~Lett.}%
\def\jcp{J.~Chem.~Phys.}%
\def\jgr{J.~Geophys.~Res.}%
\def\jqsrt{J.~Quant.~Spec.~Radiat.~Transf.}%
\def\memsai{Mem.~Soc.~Astron.~Italiana}%
\def\nphysa{Nucl.~Phys.~A}%
\def\physrep{Phys.~Rep.}%
\def\physscr{Phys.~Scr}%
\def\planss{Planet.~Space~Sci.}%
\def\procspie{Proc.~SPIED}%
\let\astap=\aap
\let\apjlett=\apjl
\let\apjsupp=\apjs
\let\applopt=\ao

\bibliographystyle{mnras}	

\bibliography{shock2d.bib}
\label{lastpage}
\end{document}